\documentclass[11pt]{article}

\usepackage{xurl}
\usepackage[hidelinks]{hyperref}
\usepackage{geometry}
\usepackage[T1]{fontenc}
\usepackage[utf8]{inputenc}
\usepackage{lineno}
\usepackage{enumitem}
\usepackage{titling}
\usepackage{etoolbox}
\usepackage{mathtools}
\usepackage{titlesec}
\usepackage{gensymb}
\usepackage{multirow}
\usepackage{placeins}
\usepackage[numbers]{natbib}
\usepackage{bibunits}
\usepackage{pdflscape}
\usepackage{makecell}
\usepackage{fancyvrb}
\usepackage{amsmath}
\usepackage{booktabs}
\usepackage{rotating}
\usepackage{graphicx}
\usepackage{adjustbox}

\defaultbibliographystyle{elsarticle-num}
\defaultbibliography{mybib}

\usepackage{tabularx}
\usepackage{threeparttable}
\usepackage[english]{babel}
\usepackage{amsfonts}
\usepackage{amssymb}
\usepackage{wasysym}
\usepackage{bbm}
\usepackage{array}
\usepackage{verbatim}
\usepackage{float}
\usepackage{caption}
\usepackage{subcaption}
\usepackage{xr}
\usepackage{breakcites}
\usepackage{longtable}
\usepackage{cleveref}
\usepackage{soul}
\usepackage{xcolor}

\titleformat{name=\section}{\normalfont\Large\bfseries}{}{0pt}{}
\titleformat{name=\subsection}{\normalfont\large\bfseries}{}{0pt}{}
\titleformat{name=\subsubsection}{\normalfont\normalsize\bfseries}{}{0pt}{}

\newcommand{\Tau}{T}

\newtagform{brackets}{[}{]}
\usetagform{brackets}

\title{Public transit gains and spatially uneven travel demand changes after NYC congestion pricing}

\newcommand{\beginsupplement}{%
  \clearpage
  \renewcommand{\thepage}{S\arabic{page}}%
  \setcounter{page}{1}%
  \setcounter{section}{0}%
  \setcounter{subsection}{0}%
  \setcounter{subsubsection}{0}%
  \setcounter{table}{0}%
  \setcounter{figure}{0}%
  \setcounter{equation}{0}%
  \renewcommand{\thesection}{S\arabic{section}}%
  \renewcommand{\thetable}{S\arabic{table}}%
  \renewcommand{\thefigure}{S\arabic{figure}}%
  \renewcommand{\theequation}{S\arabic{equation}}%
  \renewcommand{\theHsection}{supp.\arabic{section}}%
  \renewcommand{\theHsubsection}{supp.\arabic{section}.\arabic{subsection}}%
  \renewcommand{\theHsubsubsection}{supp.\arabic{section}.\arabic{subsection}.\arabic{subsubsection}}%
  \renewcommand{\theHtable}{supp.\arabic{table}}%
  \renewcommand{\theHfigure}{supp.\arabic{figure}}%
  \renewcommand{\theHequation}{supp.\arabic{equation}}%
}

\begin{document}
% ----------------------------------------------------------------------
\begin{titlepage}

{\noindent\LARGE\bf\thetitle}
\bigskip

\begin{flushleft}\large
    Donghang Li\textsuperscript{1,2,7,$\dagger$},
    Dingyi Zhuang\textsuperscript{1,$\dagger$},
    Yunlin Li\textsuperscript{3},
    Chenan Shen\textsuperscript{1,2},
    Nina Cao\textsuperscript{4},
    Yunhan Zheng\textsuperscript{5,*},
    Shenhao Wang\textsuperscript{6},
    Jinhua Zhao\textsuperscript{2}
\end{flushleft}

\bigskip
\noindent
\begin{enumerate}[label=\textbf{\arabic*}]
 \item Department of Civil and Environmental Engineering, Massachusetts Institute of Technology, Cambridge, MA, USA
 \item Department of Urban Studies and Planning, Massachusetts Institute of Technology, Cambridge, MA, USA
  \item Mathematical Institute, University of Oxford, Oxford, UK
 \item Department of Mechanical Engineering, Massachusetts Institute of Technology, Cambridge, MA, USA
 \item College of Urban and Environmental Sciences, Peking University, Beijing, China
 \item Department of Urban and Regional Planning, University of Florida, Gainesville, FL, USA
  \item Center for Computational Science and Engineering, Massachusetts Institute of Technology, Cambridge, MA, USA
\end{enumerate}

\bigskip
\textbf{$\dagger$} These authors contributed equally to this work. \\
\textbf{*} To whom correspondence should be addressed. E-mail: yunhan@pku.edu.cn
\vfill
\end{titlepage}

% ======================================================================
% ======================================================================
% ABSTRACT  ~150 words, single paragraph, no heading numbering.
% ======================================================================
\begin{bibunit}
\section{Abstract}

New York City implemented the nation’s first cordon-based congestion pricing program in January 2025, providing an opportunity to evaluate how system-wide urban mobility responds to large-scale pricing interventions. Because such policies generate spillovers across modes and locations, credible control groups are difficult to construct. We address this challenge using time series foundation models to generate probabilistic counterfactual demand forecasts with calibrated uncertainty. Applying this framework to bus, subway, and aggregate trip volume data, we find that post-policy bus and subway ridership increased significantly relative to expected no-policy demand, while overall travel demand decreased modestly. The effects are spatially heterogeneous: while reductions in overall travel demand are concentrated within the Congestion Relief Zone, transit gains extend beyond Manhattan’s core. Socio-demographic analyses further reveal uneven adaptation across neighborhoods, highlighting spatial equity implications. Our framework provides a scalable approach for the uncertainty-aware evaluation of system-wide urban interventions when clean control groups are unavailable.

\newpage

% ======================================================================
% Introduction
% Target: ~1300 words, 6 paragraphs.
% ======================================================================
\section{Introduction} 

Congestion pricing charges drivers for entering or traveling within congested urban areas, usually with the dual goal of reducing traffic externalities and funding transportation improvements \citep{vickrey1969congestion}. Economic theory and prior urban transport research suggest that such pricing can reduce inefficient peak-period driving when travelers face the social costs of congestion more directly \citep{small2013urban}. On January 5, 2025, New York City implemented the nation’s first cordon-based congestion pricing program. The policy charges vehicles entering Manhattan south of 60th Street \$9 for cars and higher rates for trucks and buses in order to reduce congestion, improve air quality, and generate stable funding for transit infrastructure \citep{mta_congestion_2025,cicbca_fhwa_approval_2024}.
Toll revenues, estimated at about \$1 billion annually, are earmarked for subway modernization, electric bus deployment, and other transit improvements. Early evidence suggests that the policy is working as intended: vehicle speeds in the Central Business District (CBD) have increased by about 15\%, and CO\textsubscript{2} emissions per kilometer have declined by 2–3\% \citep{cook2025short}. Official six-month summaries also report fewer vehicle entries, faster in-zone bus travel, and higher transit ridership compared to the previous year \citep{MTA_2025_SixMonths_CongestionPricing}.

Despite these encouraging outcomes, a fundamental question remains: are reductions in vehicle trips into the CBD driven by travelers shifting to alternative modes, or by travelers avoiding the CBD altogether \citep{eliasson2009stockholm}? This distinction is central to evaluating the broader social and economic implications of congestion pricing. If many travelers cancel trips, congestion reduction may come at the cost of accessibility and downtown activity \citep{schwach2025congestion}. Conversely, if drivers shift to public transit or other shared and lower-emission modes, the city can achieve its environmental and mobility goals without reducing participation in urban life \citep{li2010willingness}. Addressing this trade-off requires analyzing multimodal travel demand responses rather than focusing solely on vehicle traffic counts \citep{cook2025short}. Accordingly, a central empirical question is whether observed increases in public transit ridership exceed what would have been expected under normal post-pandemic recovery and seasonal dynamics. Existing evidence and agency reports suggest meaningful rising transit usage following the implementation of the congestion pricing policy. However, these descriptive comparisons are not estimated within a counterfactual framework and may conflate policy effects with underlying recovery trends, seasonality, service changes, and other contemporaneous dynamics \citep{ziedan2023will}.

New York City offers an exceptional opportunity to examine this question. As the first U.S. city to adopt congestion pricing, it features an extensive and highly interdependent transportation system whose responses extend well beyond the tolling zone \citep{jing2024evaluating}. The Congestion Relief Zone (CRZ)—Manhattan south of 60th Street—sits at the center of a dense multimodal network linking subways, buses, and regional mobility systems. Yet this interconnectedness also complicates policy evaluation: nearly all travelers across modes are affected directly or indirectly, leaving no clearly defined control group. Difference-in-differences designs require credible comparison groups with stable counterfactual trends, which are difficult to define when a policy affects a whole metropolitan network \citep{callaway2021difference}. Boundary-based designs may be complicated by rerouting, retiming, and mode switching around the pricing boundary \citep{lee2010regression,duranton2018urban}. Synthetic control methods require comparable donor units and strong pre-policy fit, which may be unavailable when spillovers extend across the candidate comparison pool \citep{abadie2010synthetic,arkhangelsky2021synthetic}.
Related evaluations of large transportation policies emphasize that spillovers and poor comparison-unit fit can make conventional quasi-experimental baselines fragile \citep{baghestani2020evaluating,morton2025impact}. As a result, conventional econometric approaches are difficult to apply credibly in this setting.

To overcome these limitations, this study introduces a probabilistic counterfactual modeling framework that replaces treated–control comparisons with explicit counterfactual forecasting. Forecasting-based counterfactual approaches have been used to construct expected outcomes when external comparison units are unavailable or difficult to justify \citep{brodersen2015inferring}. Rather than relying on single point predictions, we model the full distribution of expected travel demand based on pre-intervention patterns, allowing us to construct prediction intervals that capture normal variation arising from seasonality, weather, holidays, and other unobserved factors. Probabilistic forecasting is especially useful in this setting because uncertainty intervals distinguish ordinary temporal variation from unusually large post-policy deviations \citep{schulam2017reliable,hill2020bayesian}. Crucially, these intervals are empirically calibrated. For example, a 90\% prediction interval contains approximately 90\% of observed outcomes in pre-policy validation data. Policy impacts are inferred when post-intervention observations fall outside the expected range, providing a statistically grounded means of distinguishing genuine policy effects from routine demand fluctuations.

We operationalize this approach using TimesFM with hierarchical quantile calibration, referred to as TimesFM-HQC, which forecasts expected no-policy demand trajectories and calibrated uncertainty bounds using pre-policy data alone. Recent time series foundation models such as Chronos and TimesFM use large-scale pretraining to improve zero-shot forecasting across heterogeneous time series \citep{ansari2024chronos,das2023decoder}. Benchmarking studies show that these pretrained models can perform strongly across diverse forecasting domains, motivating their use as scalable base forecasters for transportation demand panels \citep{ahamedzero}. TimesFM-HQC produces calibrated forecasts for multiple transportation modes, including buses, subways, and overall travel, enabling consistent uncertainty-aware comparison between observed post-policy demand and expected no-policy demand in a setting where traditional control-group designs are difficult or even infeasible.

Our results indicate that congestion pricing reshaped urban mobility in New York City. While overall trip volumes declined modestly, public transit ridership increased substantially after policy implementation, with adjustment dynamics varying across modes. These effects extended beyond the congestion zone into surrounding borough corridors, revealing system-wide spillovers and distinct spatial patterns inside and outside the CRZ. Moreover, ridership responses varied systematically with neighborhood socio-economic characteristics, pointing to heterogeneous and potentially uneven impacts across communities.

This study makes four main contributions.

\begin{enumerate}
  \item It develops a \textbf{probabilistic counterfactual framework} that combines time-series foundation-model forecasting with uncertainty quantification, which provides an uncertainty-aware counterfactual forecasting approach for evaluating system-wide mobility changes when clean external controls are difficult to define.
  
  \item It provides one of the \textbf{first multimodal assessments} of New York City’s congestion pricing program, jointly examining bus ridership, subway ridership, and overall travel activity.
  
  \item It compares \textbf{public transit gains and overall travel changes}, providing evidence consistent with modal substitution playing a larger role than widespread trip suppression in the early post-policy period.
  
  \item It links estimated spatial effects to \textbf{socio-demographic characteristics}, highlighting the distributional and equity-based implications of large-scale transportation pricing policies.
\end{enumerate}

Together, these contributions demonstrate the potential of probabilistic counterfactual forecasting for evaluating large-scale urban interventions.

% ======================================================================
% RESULTS  Target ~2500 words across 5 finding-led subsections.
% ======================================================================
\section{Results}

\subsection{Probabilistic counterfactual forecasting framework.}
\label{sec:results_framework}

\begin{figure}[!t]
    \centering
    \includegraphics[width=1\linewidth]{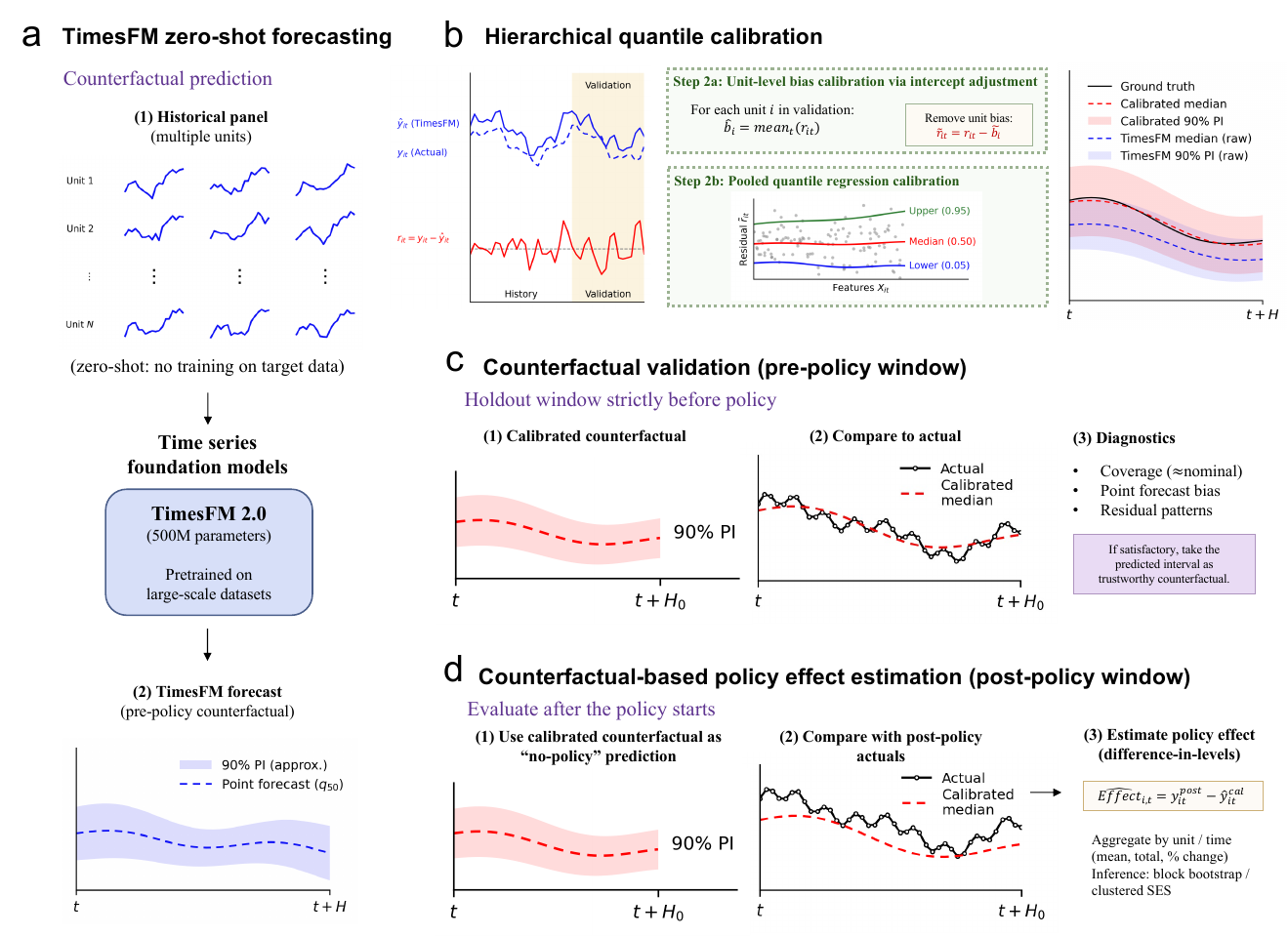}
    \caption{
    \textbf{Uncertainty-aware counterfactual forecasting framework.}
    \textbf{a,} a pretrained time series foundation model (TimesFM 2.0) generates zero-shot probabilistic forecasts from historical panel data, producing point forecasts and native quantile outputs that are used to construct prediction intervals.
    \textbf{b,} forecast uncertainty is calibrated using validation residuals through a hierarchical quantile calibration (HQC) framework. Residuals from the validation window are modeled using calendar and demand-level features, combining unit-specific bias correction with pooled quantile regression to improve interval calibration and robustness across heterogeneous units.
    \textbf{c,} calibrated forecasts are evaluated on a held-out pre-policy validation window to assess predictive accuracy, interval coverage, and residual behavior before post-policy comparison.
    \textbf{d,} the calibrated post-policy forecasts are used as counterfactual ``no-policy'' trajectories and compared with observed post-policy demand to estimate changes relative to expected demand over time and across travel modes.
    }
    \label{fig:pic1}
\end{figure}

Figure~\ref{fig:pic1} presents our probabilistic counterfactual forecasting framework for measuring post-policy mobility changes relative to expected no-policy demand. Because congestion pricing may generate system-wide spillovers across locations, routes, and travel modes, conventional control groups are difficult to define. Instead, we construct a model-based counterfactual representing the demand trajectory that would have been expected in the absence of the policy.

The framework consists of three steps. First, TimesFM 2.0 generates zero-shot forecasts from historical demand data. For each spatial or modal unit, the model produces both point forecasts and quantile predictions over the post-policy horizon, providing a scalable forecasting backbone without requiring separate model training for each unit. Second, we calibrate forecast uncertainty using residuals from a held-out pre-policy validation window. This hierarchical quantile calibration corrects persistent unit-level bias and improves the reliability of prediction intervals. Detailed calibration procedures are provided in Methods and Supplementary Information. Third, we validate the calibrated forecasts before post-policy comparison. Because the validation period precedes policy implementation, observed demand provides a direct benchmark for evaluating forecast accuracy, interval coverage, and residual behavior. 

In the post-policy period, calibrated forecasts are interpreted as no-policy expected demand trajectories. We then compare observed demand with these calibrated forecasts and aggregate the resulting deviations across time, spatial units, and travel modes. This framework therefore combines the flexibility of time series foundation models with calibration-based uncertainty quantification, enabling the statistically grounded measurement of demand changes in a complex multimodal transportation system.

% ----------------------------------------------------------------------
\subsection{Public transit demand increased while overall travel changed modestly}
\label{sec:estimation_results}

\begin{figure}[!t]
    \centering
    \includegraphics[width=0.9\linewidth]{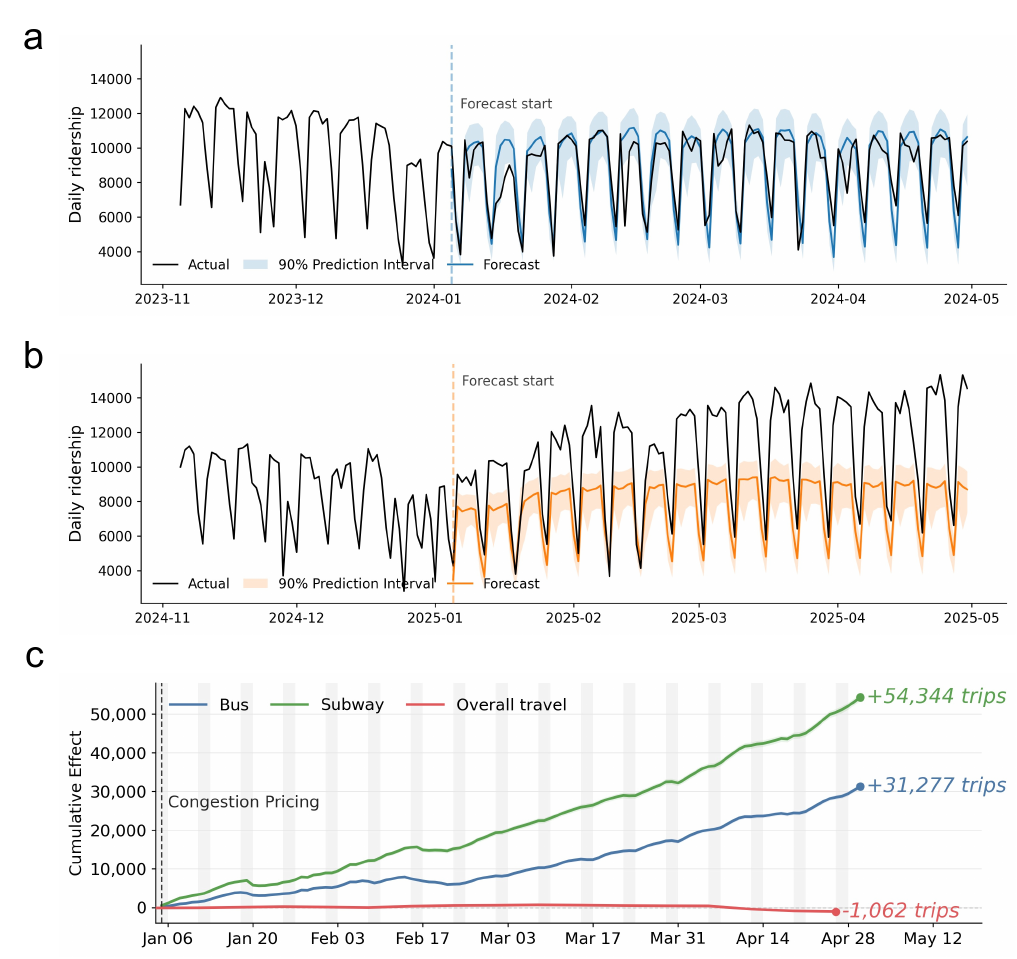}
    \caption{
    \textbf{Probabilistic counterfactual forecasts and cumulative changes after congestion pricing.}
    \textbf{a,} Example validation-period forecast for a representative bus route. The black line shows observed daily ridership, the blue line shows the model forecast, and the shaded band denotes the 90\% prediction interval. The vertical dashed line marks the start of the forecast window. This panel illustrates the model's ability to reproduce pre-policy temporal patterns under a held-out validation setting.
    \textbf{b,} Example post-policy counterfactual forecast for the same representative series. The black line shows observed ridership after the implementation of congestion pricing, while the orange line and shaded band show the predicted counterfactual demand and its 90\% prediction interval in the absence of the policy.
    \textbf{c,} Estimated cumulative deviations from expected demand per spatial unit after the start of congestion pricing. Lines show cumulative per-unit changes for bus ridership, subway ridership, and overall travel demand measured from Replica data, with shaded vertical bands indicating weekends. Values at the end of each line report the final cumulative per-unit change over the study period.
    }
    \label{fig:pic2}
\end{figure}

As shown in Fig.~\ref{fig:pic2}, the proposed probabilistic counterfactual framework provides accurate and well-calibrated forecasts for multimodal travel demand. During the pre-policy validation period, the model successfully reproduces observed demand dynamics, including weekly seasonality and recurrent fluctuations, with observed values generally falling within the 90\% prediction interval (Fig.~\ref{fig:pic2}a). This forecasting performance supports the use of calibrated forecasts as no-policy counterfactual trajectories.

Following the implementation of congestion pricing on January 5, 2025, systematic deviations emerge between observed demand and counterfactual forecasts. Figure~\ref{fig:pic2}b illustrates a representative bus route for which observed ridership consistently exceeds expected no-policy demand. Similar patterns are observed across the transit system, particularly for subway ridership.

Figure~\ref{fig:pic2}c summarizes cumulative demand changes over the study period. Bus ridership increased by approximately 31,277 rides per route, corresponding to roughly 7.7 million additional rides across 246 routes. Subway ridership increased by approximately 54,344 rides per station, equivalent to about 23.0 million additional rides across 424 stations. In contrast, overall travel demand decreased by approximately 1,062 trips per census tract, corresponding to about 2.08 million fewer trips across the city.

The increase in public transit demand substantially exceeds the reduction in overall travel demand, suggesting that modal substitution played a larger role than widespread trip suppression during the early post-policy period. Subway gains were larger than bus gains, indicating that rail transit absorbed a substantial share of adjusted travel demand.

These estimates are broadly consistent with MTA reports documenting approximately 7\% growth in subway ridership and 12\% growth in bus ridership following congestion pricing \citep{mta_congestion_2025}. Our framework estimates a 6.04\% increase in subway ridership and a 6.4\% increase in bus ridership relative to expected no-policy demand, yielding directionally consistent but more conservative estimates after accounting for underlying temporal trends and seasonality.

% ----------------------------------------------------------------------
\subsection{Demand changes vary across the CRZ boundary}
\label{sec:results_crz}

\begin{figure}[!t]
    \centering
    \includegraphics[width=0.9\linewidth]{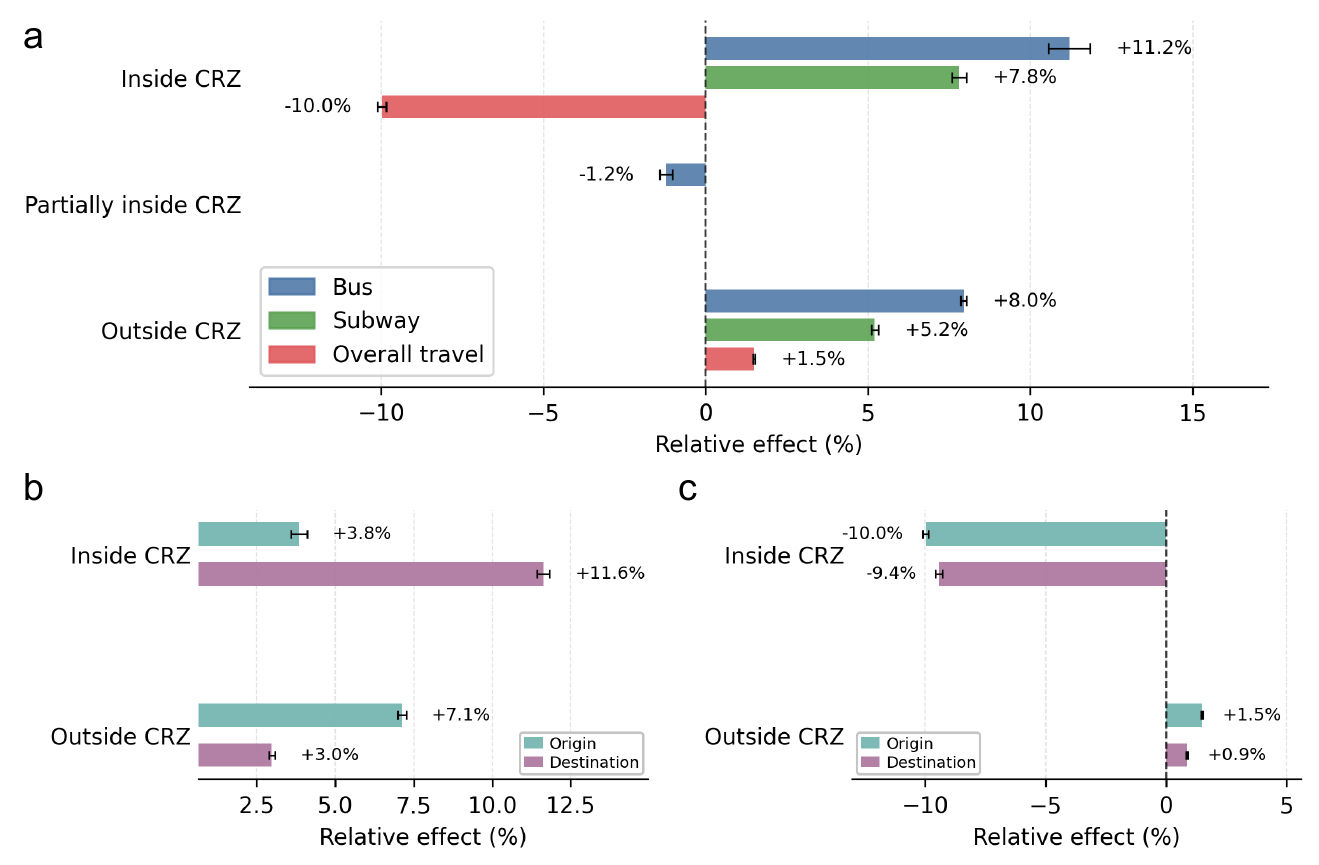}
    \caption{
    \textbf{CRZ-specific changes relative to expected demand across travel modes and trip directions.}
    \textbf{a,} Relative changes within and outside the Congestion Relief Zone (CRZ) across bus, subway, and overall travel demand. Bus ridership increases both inside and outside the CRZ, while overall travel demand decreases substantially inside the CRZ but increases slightly outside the CRZ. Subway ridership also shows positive changes, consistent with increased public transit use after congestion pricing implementation.
    \textbf{b,} Relative changes for subway entries and exits. Both origin and destination subway demand increase inside and outside the CRZ, with stronger destination-side changes observed inside the CRZ.
    \textbf{c,} Relative changes for overall travel origins and destinations. Overall travel demand declines markedly inside the CRZ for both origins and destinations, while modest increases are observed outside the CRZ, indicating spatial redistribution of travel demand beyond the tolling zone. Error bars represent 95\% confidence intervals.
    }
    \label{fig:pic3}
\end{figure}

Post-policy demand changes exhibit substantial spatial heterogeneity across the Congestion Relief Zone (CRZ) boundary. Figure~\ref{fig:pic3}a shows that public transit demand increases both inside and outside the CRZ, whereas overall travel demand declines sharply inside the tolling zone. Specifically, bus ridership increases by approximately 11.2\% inside the CRZ and 8.0\% outside the CRZ, while subway ridership increases by 7.8\% inside the CRZ and 5.2\% outside the CRZ. In contrast, overall travel demand measured using Replica data decreases by nearly 10\% inside the CRZ, but slightly increases outside the CRZ. These patterns are consistent with reduced vehicle-based travel within the tolling area and increased public transit use across the broader metropolitan region.

The spatial redistribution pattern becomes clearer when separating subway origin and destination flows. As shown in Figure~\ref{fig:pic3}b, subway entries outside the CRZ increase by approximately 7.1\%, while subway exits inside the CRZ increase substantially by 11.6\%. In comparison, subway entries inside the CRZ increase by only 3.8\%, and subway exits outside the CRZ increase by approximately 3.0\%. This asymmetric pattern is consistent with increased subway-based access from outside the tolling zone into the CRZ after congestion pricing implementation. In other words, the strongest subway change is concentrated on destination demand within the CRZ, reflecting increased transit-based access to the urban core.

A similar but opposite spatial pattern is observed for overall travel demand. Figure~\ref{fig:pic3}c shows that both travel origins and destinations decline substantially inside the CRZ, with relative changes of approximately $-10.0\%$ and $-9.4\%$, respectively. Outside the CRZ, however, overall travel demand increases slightly, with positive relative changes below 2\%. Together, these results indicate that post-policy demand changes are not uniform across the city. Instead, observed demand shifts away from the congested urban core toward surrounding regions while public transit use increases for accessing the CRZ.

% ----------------------------------------------------------------------
\subsection{Spatial heterogeneity in post-policy demand changes}
\label{sec:results_spatial}

\begin{figure}[!t]
    \centering
    \includegraphics[width=1\linewidth]{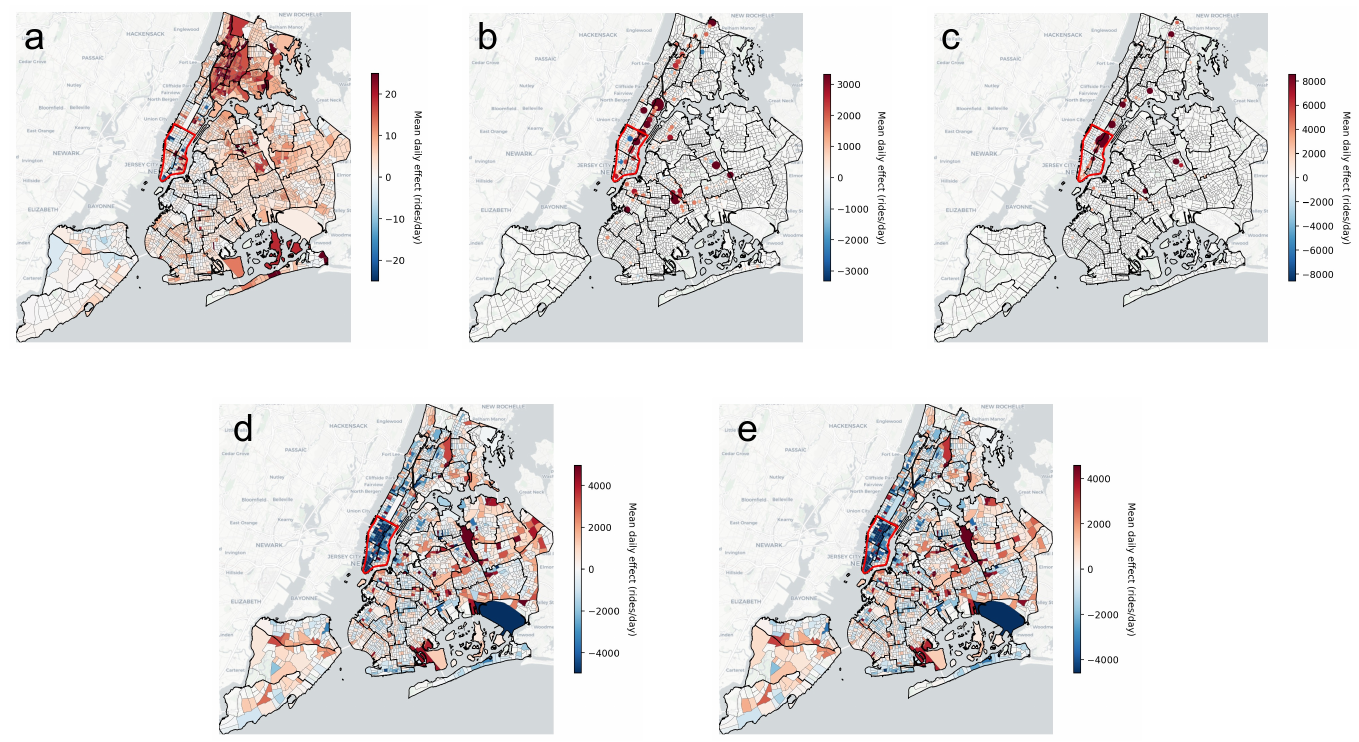}
    \caption{
    \textbf{Spatial distribution of estimated changes relative to expected demand across different transportation modes.}
    Each panel presents the average post-policy change relative to the calibrated counterfactual baseline at the census tract level after the implementation of congestion pricing in New York City. Positive values indicate increased demand relative to the counterfactual baseline, while negative values indicate demand reductions. The red polygon outlines the Congestion Relief Zone (CRZ).
    \textbf{a,} Bus ridership changes, showing substantial increases especially around CRZ areas, suggesting increased bus use.
    \textbf{b,} Subway origin ridership changes, highlighting heterogeneous increases concentrated near major transit hubs.
    \textbf{c,} Subway destination ridership changes, with strong positive changes in central Manhattan.
    \textbf{d,} Overall travel origin changes, showing widespread reductions in trip generation, particularly within the CRZ.
    \textbf{e,} Overall travel destination changes, indicating reduced inbound travel demand toward Manhattan alongside localized increases in outer areas.
    Together, these results reveal that congestion pricing reshapes multimodal travel demand with strong spatial heterogeneity across New York City.
    }
    \label{fig:pic4}
\end{figure}

Figure~\ref{fig:pic4} further illustrates the strong spatial heterogeneity of post-policy demand changes across New York City at the census-tract level. Rather than producing uniform patterns, congestion pricing coincides with highly localized changes in multimodal travel demand, with distinct spatial patterns across bus, subway, and overall travel activities.

Bus ridership changes are positive throughout much of the city, with the strongest increases concentrated around Manhattan and neighborhoods adjacent to the Congestion Relief Zone (CRZ), as shown in Figure~\ref{fig:pic4}a. This pattern suggests that bus services may have served as an important substitute for short- and medium-distance vehicle trips into the urban core after the implementation of congestion pricing. Several outer-borough areas also experience moderate increases, indicating that post-policy changes extend beyond the immediate tolling zone through network-wide modal adjustments.

Subway changes display a more concentrated and hub-oriented spatial structure. Figure~\ref{fig:pic4}b shows that subway origin demand increases are primarily clustered around major residential and transfer areas outside the CRZ, indicating increased subway boarding activity from peripheral neighborhoods. In contrast, Figure~\ref{fig:pic4}c reveals that subway destination changes are strongly concentrated within central Manhattan, particularly inside and near the CRZ boundary. This asymmetric spatial pattern is consistent with increased transit-based access into the Manhattan core.

The spatial distribution of overall travel demand presents a substantially different pattern. Figures~\ref{fig:pic4}d and~\ref{fig:pic4}e show widespread reductions in both trip origins and destinations within the CRZ, consistent with the direct disincentive effect of congestion pricing on vehicle-based travel. The strongest negative effects are concentrated in lower Manhattan and nearby high-density corridors, where congestion charges are directly imposed. Outside the CRZ, however, the effects become more heterogeneous, with localized increases observed in parts of the outer boroughs. This suggests that some travel demand is spatially redistributed rather than entirely eliminated, potentially reflecting rerouting, relocation of activities, or induced travel toward areas outside the tolling zone.

Overall, the spatial analysis shows that post-policy demand changes are highly uneven geographically. Public transit gains are concentrated near major commuting corridors and central employment areas, while reductions in overall travel demand are primarily localized within the CRZ. These findings highlight the importance of considering spatial heterogeneity, as aggregate citywide averages may obscure substantial neighborhood-level differences in accessibility and mobility adaptation.

% ---------------------------------------------------------------------

\subsection{Socio-demographic correlates of spatial effects}

\begin{figure}[!t]
    \centering
    \includegraphics[width=1\linewidth]{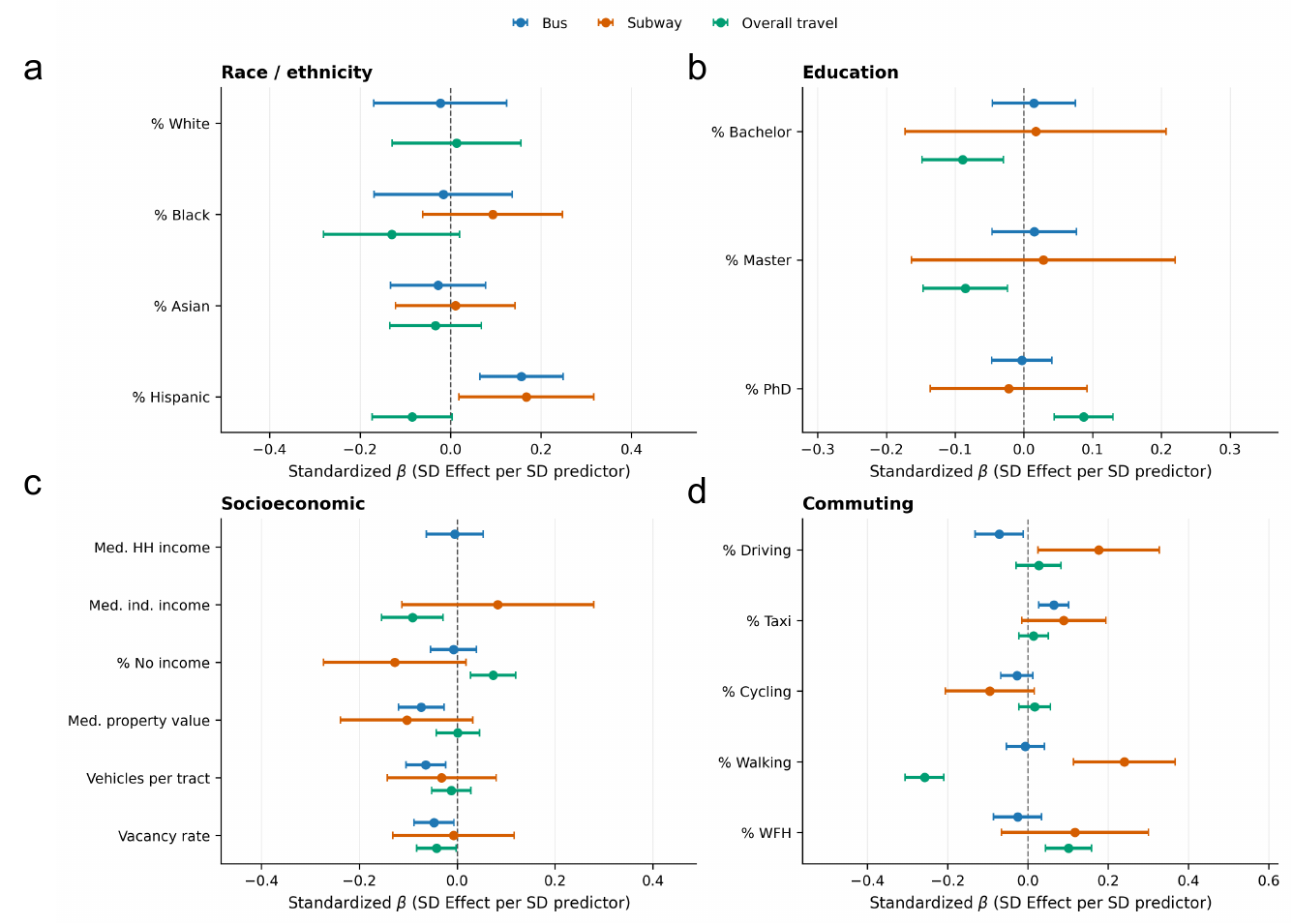}
    \caption{
    \textbf{Socio-demographic correlates of spatial effects across transportation modes.}
    Standardized regression coefficients ($\beta$) and corresponding 90\% confidence intervals are shown for associations between census-tract socio-demographic characteristics and estimated average effects of congestion pricing. Positive coefficients indicate that areas with higher values of the corresponding predictor experienced larger increases in demand relative to the counterfactual baseline, while negative coefficients indicate larger reductions. Blue, orange, and green markers correspond to bus, subway, and overall travel effects, respectively.
    \textbf{a,} Race and ethnicity characteristics.
    \textbf{b,} Educational attainment characteristics.
    \textbf{c,} Socioeconomic characteristics.
    \textbf{d,} Commuting behavior characteristics.
    Overall, the results reveal potential spatial inequality in congestion pricing impacts, with heterogeneous responses across demographic and socioeconomic groups as well as transportation modes.
    }
    \label{fig:pic5}
\end{figure}

Figure~\ref{fig:pic5} examines whether spatially heterogeneous demand changes are associated with local socio-demographic characteristics. These estimates are descriptive rather than causal and are intended to characterize which types of neighborhoods experienced larger or smaller responses to congestion pricing.

Panel~a shows substantial heterogeneity across race and ethnicity groups. Census tracts with higher Hispanic population shares are associated with larger increases in bus and subway ridership but larger reductions in overall travel demand. In contrast, coefficients for White and Asian population shares are generally weaker and closer to zero. This pattern may suggest that some communities experienced stronger shifts toward public transit while also seeing reductions in overall travel demand.

Educational attainment also exhibits heterogeneous associations (Panel~b). Areas with higher shares of residents holding bachelor’s or master’s degrees tend to experience larger reductions in overall trips, whereas areas with higher shares of doctoral degree holders are associated with more positive trip changes.

Socioeconomic characteristics are strongly associated with mobility responses (Panel~c). Median individual income is negatively associated with overall travel effects, suggesting larger trip reductions in higher-income neighborhoods. Given the concentration of higher-income census tracts in Manhattan and adjacent core areas, this pattern indicates that travel-demand reductions are disproportionately concentrated in the urban core. Vehicle ownership and median property value are negatively associated with bus effects, implying weaker public transit gains in more car-dependent and higher-value neighborhoods. This heterogeneity raises equity concerns, as the behavioral response to pricing is shaped not only by the toll itself, but also by the availability of substitutes and the capacity of different communities to adjust.

Finally, pre-existing commuting behavior is also associated with heterogeneous responses (Panel~d). Areas with higher pre-policy driving shares exhibit stronger subway gains but weaker bus effects, while work-from-home shares are positively associated with both subway and overall travel changes. Together, these results suggest that mobility responses to congestion pricing vary systematically across neighborhoods and are closely related to socioeconomic conditions, travel behavior, and access to alternative transportation options.

% ======================================================================
% DISCUSSION  Target ~1100 words, 5--6 paragraphs.
% ======================================================================
\section{Discussion}

% new

New York City’s congestion pricing program, implemented on January 5, 2025, represents the first cordon-based pricing policy in the United States and provides a unique opportunity to examine how large-scale pricing interventions reshape multimodal urban mobility. A central policy question is whether congestion pricing primarily suppresses overall travel demand or instead reallocates trips toward more sustainable transportation modes. This distinction is critical because the welfare and accessibility implications of congestion pricing differ substantially depending on whether travelers reduce mobility altogether or adapt through modal substitution. Prior congestion-pricing research similarly emphasizes that traffic reduction, public transport substitution, and accessibility changes may imply fundamentally different welfare interpretations \citep{small2013urban,eliasson2008lessons}.

Existing articles and reports have provided valuable descriptive evidence on early congestion pricing outcomes \citep{mta2025icymi}. Official monitoring has similarly emphasized changes in traffic, transit, and travel speeds during the first months of implementation \citep{MTA_2025_SixMonths_CongestionPricing}. However, simple before--after comparisons may be confounded by seasonality, recovery dynamics, service changes, holidays, and other contemporaneous shocks. Therefore, recent studies have increasingly emphasized counterfactual construction to isolate policy impacts from other time-varying influences \citep{li2026quantifying}. Our contribution is to place these early patterns within an uncertainty-aware counterfactual forecasting framework that estimates what demand would have been expected in the absence of the policy, based only on pre-policy temporal data.  

Empirically, our results are consistent with modal substitution playing a larger role than widespread trip suppression in the early post-policy period, although both effects do exist. Relative to calibrated no-policy expectations, subway ridership increased by approximately 6.0\% and bus ridership increased by approximately 6.4\%, corresponding to an estimated 23.0 million additional subway rides and 7.7 million additional bus rides during the study period. These findings are directionally aligned with official MTA reports while offering a more conservative comparison that accounts for temporal trends, seasonality, and latent demand dynamics. In contrast, overall travel demand measured using Replica data declined, but by a much more modest amount, corresponding to approximately 2.08 million fewer trips citywide. These patterns suggest that congestion pricing primarily reshaped how people travel rather than substantially reducing urban mobility itself.

The estimated demand changes also display strong spatial heterogeneity. Public transit gains extend beyond the Congestion Relief Zone (CRZ) into surrounding boroughs, indicating that congestion pricing coincides with system-wide mobility adjustments rather than purely localized changes. Bus ridership increased by approximately 11.2\% inside the CRZ and 8.0\% outside the CRZ, while subway ridership increased by 7.8\% and 5.2\%, respectively. In contrast, overall travel demand declined by nearly 10\% inside the CRZ, especially within lower Manhattan and adjacent high-density corridors, while changing only modestly outside the tolling zone. The particularly strong increase in subway destination demand inside the CRZ is consistent with greater transit-based access to Manhattan after policy implementation.

Beyond aggregate mobility changes, our exploratory sociodemographic analysis shows that spatial demand responses are systematically correlated with neighborhood characteristics. This finding underscores the spatial and social equity implications of large-scale transportation policies and motivates future distributional analyses that link mode-specific demand changes to accessibility, travel-time reliability, household constraints, and welfare outcomes. Prior studies on congestion pricing similarly emphasize that public acceptance and distributional interpretation depend on how policy costs, benefits, and revenue reinvestment are allocated across travelers \citep{eliasson2014stockholm,li2010willingness}.

Methodologically, this study demonstrates the value of combining time series foundation models with uncertainty quantification for urban counterfactual forecasting. Unlike traditional approaches that rely on explicit control groups, the proposed framework enables scalable counterfactual evaluation under complex system-wide interventions where spillovers invalidate standard identification assumptions. This framework is potentially applicable to a broad range of urban policy settings, including transit network redesign, infrastructure disruption, environmental regulation, and large-scale pricing intervention. More broadly, this study illustrates how foundation models may enable a new class of policy evaluation frameworks for large-scale urban interventions where conventional quasi-experimental designs are difficult to implement.

Several limitations and future directions remain. First, forecast-based counterfactuals cannot fully rule out contemporaneous shocks that are not captured by pre-policy temporal patterns or calibration features. Second, aggregate datasets cannot directly identify individual mode switching, departure-time adjustment, rerouting behavior, or trip cancellation. Third, public transit ridership and Replica-based general travel demand differ in measurement units, coverage, and spatial resolution, so their magnitudes should be compared with caution. Fourth, the current analysis focuses on short-term responses before April 30, 2025; longer-term evaluation is needed to determine whether the observed changes persist, attenuate, or evolve alongside continued adaptation and network investment.

In summary, our findings indicate that NYC congestion pricing was followed by substantial increases in public transit usage relative to calibrated no-policy expectations, alongside more modest reductions in aggregate travel demand. These patterns are consistent with modal substitution playing a larger role than widespread trip suppression, while spatial heterogeneity and sociodemographic correlations motivate future distributional research. The proposed TimesFM-HQC framework provides a robust and uncertainty-aware approach for studying complex transportation interventions under system-wide spillovers. Rather than simply discouraging mobility, the early effects of NYC congestion pricing appear to reflect a broader reorganization of urban travel behavior toward public transportation. 
% ======================================================================
% METHODS  Target ~2500 words, 5 subsections.
% ======================================================================
\section{Methods}

\subsection{Data}
\label{sec:methods_data}

We examine mobility changes after New York City's congestion pricing program using three complementary data sources: MTA bus ridership, MTA subway ridership, and Replica trip counts. These datasets capture public transit demand and broader origin--destination travel patterns at different spatial resolutions. Detailed descriptions of the data sources, native resolutions, preprocessing procedures, and summary tables are provided in the Supplementary Information.

In the main analysis, bus ridership is measured at the route level using MTA hourly boarding records aggregated to daily demand. Subway ridership is measured at the station level using MTA hourly entries and exits aggregated to daily demand. Overall travel is measured using Replica origin--destination trip counts at the census-tract level and aggregated to weekly flows. Together, these outcomes allow us to examine how congestion pricing affected public transit use and broader travel demand across modes and space.

All counterfactual modeling and post-policy comparisons use observations before April 30, 2025. We focus on this short-term post-policy window because early responses are closer to the implementation date, whereas longer-horizon estimates may be affected by behavioral adaptation and increasing forecasting uncertainty.

\subsection{Probabilistic counterfactual framework}
\label{sec:main_framework}

We study New York City's congestion pricing program using a probabilistic counterfactual forecasting framework. The central challenge is that the policy was implemented at the scale of an interconnected urban transportation system, making it difficult to define an unaffected control group. Instead of relying on external comparison units, we estimate the travel demand that would have been expected in the absence of congestion pricing and compare this counterfactual trajectory with observed post-policy demand.
This forecasting-based design builds on earlier counterfactual time-series approaches while adapting them to a high-dimensional multimodal transportation setting \citep{brodersen2015inferring,hill2020bayesian}.

Let $y_{i,t}$ denote observed travel demand for spatial unit $i$ at time $t$, where units correspond to bus routes, subway stations, or census tracts depending on the dataset. Let $t_0$ denote the implementation date of congestion pricing. For each unit, we use only pre-policy observations,
\begin{equation}
  \mathbf{y}_{i,1:t_0-1}
  =
  \{y_{i,1},y_{i,2},\dots,y_{i,t_0-1}\},
\end{equation}
to estimate a no-policy counterfactual forecast for the post-policy period. The post-policy change relative to expected demand is defined as the deviation between observed demand and the calibrated counterfactual prediction:
\begin{equation}
  \tau_{i,t}
  =
  y_{i,t} - \hat{y}^{c}_{i,t},
  \quad t \geq t_0,
  \label{eq:main_tau}
\end{equation}
where $\hat{y}^{c}_{i,t}$ is the calibrated counterfactual demand. Positive values indicate observed demand above the no-policy baseline, whereas negative values indicate observed demand below the no-policy baseline.

The framework consists of three steps: base counterfactual forecasting, validation-based calibration, and post-policy comparison. Figure~\ref{fig:pic1} summarizes the workflow.

\subsubsection{Base counterfactual forecasting with time series foundation models}
\label{sec:main_base_forecasting}

We use time series foundation models as the base forecasters. These models are pretrained on large-scale heterogeneous time series and can generate zero-shot forecasts without task-specific model training. For each unit $i$, the base model produces an initial counterfactual prediction:
\begin{equation}
  \mu_{i,t}
  =
  f_{\theta}\left(\mathbf{y}_{i,1:t_0-1}\right),
  \quad t \geq t_0,
  \label{eq:main_base_forecast}
\end{equation}
where $\mu_{i,t}$ denotes the raw no-policy forecast. In the main specification, we use TimesFM as the primary base model and report Chronos-based estimates as robustness checks.
TimesFM uses a decoder-only architecture trained for general-purpose time-series forecasting \citep{das2023decoder}.
Chronos instead tokenizes numerical time series and trains language-model architectures for probabilistic forecasting, providing a useful robustness benchmark \citep{ansari2024chronos}.
Recent cross-domain evaluations show that pretrained time-series models can generalize well across forecasting tasks, though local calibration remains important for domain-specific applications \citep{ahamedzero}.

Although foundation models provide scalable forecasts, their raw predictions may exhibit systematic bias when applied to local transportation demand data. For example, some routes or stations may be consistently over- or under-predicted, and raw prediction intervals may be miscalibrated for specific ridership patterns. We therefore introduce a validation-based calibration layer before using the forecasts for post-policy comparison.

\subsubsection{Hierarchical quantile calibration}
\label{sec:main_hqc}

To correct both point forecast bias and interval miscalibration, we use hierarchical quantile calibration (HQC). The key idea is to separate persistent unit-level bias from shared uncertainty patterns across the panel.
The first component is a location calibration step: each unit receives a residual intercept that shifts the raw forecast toward the empirical center of its validation errors.
This mirrors post-hoc forecast recalibration methods that estimate bias-corrected forecast centers from historical forecast--observation pairs before forming predictive distributions \citep{raftery2005bma,siegert2016recalibration}.
The second component is a quantile calibration step, which targets the residual distribution around the corrected forecast center and therefore calibrates prediction intervals rather than only average forecast error \citep{gneiting2005emos,romano2019cqr}.
Because travel demand exhibits strong calendar dependence and heterogeneous variability across routes, stations, and tracts, covariate-dependent residual quantiles are preferable to a single global residual threshold.
This is consistent with work on adaptive conformal inference and context-aware time-series calibration, which shows that predictive uncertainty may need to adjust as residual distributions vary across time or operating conditions \citep{gibbs2021adaptive,chen2023calibration}.
Related transportation applications have similarly emphasized distribution-aware and sparsity-aware uncertainty calibration for zero-inflated or long-tailed spatiotemporal demand outcomes \citep{jiang2023tweedie,zhuang2024sauc}.

We first split the pre-policy period into a model context window and a validation window. For each validation observation, we compute the forecast residual:
\begin{equation}
  r_{i,t}
  =
  y_{i,t} - \mu_{i,t}.
  \label{eq:main_residual}
\end{equation}
These residuals measure how the raw foundation model forecast deviates from observed demand before the policy intervention.

HQC proceeds in two stages. First, we estimate a unit-specific residual intercept:
\begin{equation}
  b_i
  =
  \frac{1}{|\mathcal{V}_i|}
  \sum_{t \in \mathcal{V}_i}
  r_{i,t},
  \label{eq:main_intercept}
\end{equation}
where $\mathcal{V}_i$ denotes the validation observations available for unit $i$. This intercept captures persistent bias for each route, station, or census tract.
Using a unit-specific intercept is important because raw foundation-model forecasts can be systematically too high or too low for particular routes, stations, or census tracts even when their average performance is strong across the full panel.

Second, after removing the unit-specific intercept, we fit pooled quantile regression models to the de-biased residuals:
\begin{equation}
  Q_q\left(r_{i,t}-b_i \mid \mathbf{x}_{i,t}\right)
  =
  \mathbf{x}_{i,t}^{\top}\boldsymbol{\beta}_q,
  \quad q \in \{0.05,0.50,0.95\}.
  \label{eq:main_qr}
\end{equation}
The feature vector $\mathbf{x}_{i,t}$ includes calendar and demand-level predictors, such as day-of-week indicators, month indicators, federal holiday indicators, raw forecast levels, historical demand levels, and interactions between day-of-week and demand level.

For each post-policy observation, the calibrated residual quantile is:
\begin{equation}
  \hat{d}^{q}_{i,t}
  =
  b_i + \mathbf{x}_{i,t}^{\top}\hat{\boldsymbol{\beta}}_q.
  \label{eq:main_delta}
\end{equation}
The calibrated counterfactual median and the nominal 90\% prediction interval are then given by:
\begin{align}
  \hat{y}^{c}_{i,t}
  &=
  \mu_{i,t} + \hat{d}^{0.50}_{i,t},
  \label{eq:main_calibrated_mean} \\
  \left[
  \hat{y}^{L,0.90}_{i,t},
  \hat{y}^{U,0.90}_{i,t}
  \right]
  &=
  \left[
  \mu_{i,t} + \hat{d}^{0.05}_{i,t},
  \mu_{i,t} + \hat{d}^{0.95}_{i,t}
  \right].
  \label{eq:main_calibrated_interval}
\end{align}

This design is motivated by the small-sample, multi-unit structure of transportation panel data. A fully unit-specific quantile calibration model may overfit because each unit has limited validation observations. In contrast, HQC uses one intercept per unit to correct persistent level bias while learning the prediction interval shape from the full panel. Pooling information across units follows the same practical motivation as global-local forecasting and panel-learning approaches: unit-level idiosyncrasies are retained while common error structure is estimated from a larger sample \citep{salinas2020deepar,montero2021locality}. Simulation evidence on global forecasting models further shows that pooled models can be competitive when many related series are short, heterogeneous, or only partially informative on their own \citep{hewamalage2022global}.

For comparison, we also consider a global quantile regression (QR) calibration baseline that directly fits residual quantiles using pooled validation errors without unit-specific intercept correction. As shown in Supplementary Information, the global QR baseline reduces average interval miscalibration but does not adequately remove spatially structured forecast bias, whereas HQC substantially attenuates such bias by explicitly correcting persistent unit-level deviations.

\subsubsection{Validation and post-policy comparison}
\label{sec:main_validation_effect}

Before estimating post-policy deviations, we evaluate counterfactual credibility using a held-out pre-policy validation window. We assess point prediction accuracy using RMSE, MAE, and SMAPE, and evaluate interval calibration using the empirical coverage rate of the nominal 90\% prediction interval. A reliable counterfactual model should produce accurate point forecasts and prediction intervals whose empirical coverage is close to the nominal level.
This validation logic follows standard forecasting practice, where out-of-sample accuracy and calibrated uncertainty are evaluated before interpreting forecasts in downstream analysis \citep{box2015time,taylor2018forecasting}.

After validation, the calibrated post-policy forecasts are interpreted as no-policy counterfactual trajectories. Pointwise deviations from expected demand are estimated using Eq.~\eqref{eq:main_tau}. We classify a unit-time deviation as statistically distinguishable from expected demand when the observed value falls outside the calibrated counterfactual interval:
\begin{equation}
  y_{i,t}
  \notin
  \left[
  \hat{y}^{L,0.90}_{i,t},
  \hat{y}^{U,0.90}_{i,t}
  \right].
  \label{eq:main_significance}
\end{equation}

We summarize deviations across time and space using cumulative changes, average changes, relative changes, and significance shares. These summaries are computed for each travel mode and for spatial strata defined by their relationship to the Congestion Relief Zone (CRZ), including units inside, outside, or partially overlapping the CRZ when applicable. Additional details on metric definitions, uncertainty aggregation, and robustness checks are provided in Supplementary Methods.

\subsection{Reporting summary}

Further information on research design is available in the Nature Portfolio Reporting Summary linked to this article.

% ======================================================================
% DATA / CODE  AVAILABILITY
% ======================================================================
\vspace{2mm}
\section*{Data availability}

MTA Bus Hourly Ridership data are publicly available at
\url{https://catalog.data.gov/dataset/mta-bus-hourly-ridership-beginning-2025}.
MTA Subway Hourly Ridership and Origin-Destination Estimates are
available at
\url{https://data.ny.gov/Transportation/MTA-Subway-Hourly-Ridership-2020-2024/wujg-7c2s}
and
\url{https://www.mta.info/article/introducing-subway-origin-destination-ridership-dataset}.
Replica aggregate trip counts data are from \url{https://www.replicahq.com/} through subscription.
U.S. Census tract-level demographics are obtained from the 2020 American Community Survey (\url{https://www.census.gov/programs-surveys/acs}). Geographic delineations of the Congestion Relief Zone are obtained from the Metropolitan Transportation Authority
(\url{https://congestionreliefzone.mta.info}). 

\vspace{2mm}
\section*{Code availability}

The analysis was conducted using Python. The scripts that support the findings of this study are also available via the GitHub repository at
\url{https://github.com/MrDonghang/nyc_congestion_pricing.git}.
% ======================================================================
% REFERENCES
% ======================================================================
\begingroup
\renewcommand{\refname}{References}
\putbib
\endgroup

% ======================================================================
% ACKNOWLEDGEMENTS, AUTHOR CONTRIBUTIONS, COMPETING INTERESTS
% ======================================================================
\section*{Acknowledgements}

This research is supported by the National Research Foundation (NRF), Prime Minister’s Office, Singapore under its Campus for Research Excellence and Technological Enterprise (CREATE) programme. The Mens, Manus, and Machina (M3S) is an interdisciplinary research group (IRG) of the Singapore MIT Alliance for Research and Technology (SMART) centre.

\section*{Author Contributions Statement}

Donghang Li: Writing – review \& editing, Writing – original draft, Visualization, Methodology, Investigation, Formal analysis, Data curation, Conceptualization; Dingyi Zhuang: Writing – review \& editing, Writing – original draft, Visualization, Methodology, Investigation, Formal analysis, Data curation, Conceptualization; Yunlin Li: Writing – review \& editing, Data analysis; Chenan Shen: Writing – review \& editing, Data analysis; Yunhan Zheng: Writing – review \& editing, Supervision, Methodology, Formal analysis, Conceptualization; Shenhao Wang: Supervision, Project administration; Jinhua Zhao: Supervision, Project administration, Funding acquisition.

\section*{Competing Interests Statement}

The authors declare no competing interests.
\end{bibunit}

% ======================================================================
% SUPPLEMENTARY INFORMATION
% ======================================================================
\beginsupplement
\begin{bibunit}
\section*{Supplementary Information}
\addcontentsline{toc}{section}{Supplementary Information}
\tableofcontents

\section{Extended literature review}
\label{sec:si_litreview}

This section provides an extended treatment of three bodies of literature
that motivate our approach: cordon-based congestion pricing in
international practice; quasi-experimental approaches in
urban transportation policy; and forecasting-based counterfactual
analysis for systemwide interventions.

\subsection{International experience with cordon pricing}
\label{sec:si_litreview_global}

Congestion pricing is a policy that charges drivers for entering or traveling within congested urban areas, usually central business districts \citep{lindsney2001traffic}.
The goal is to reduce traffic and emissions while generating revenue for public transit improvements \citep{small2013urban,eliasson2014stockholm}.
By raising the cost of peak-hour driving, congestion pricing encourages travelers to switch to alternative modes, adjust their trip timing, or forgo trips altogether \citep{li2010willingness}.

Over the past three decades, six major cities have implemented full-scale congestion pricing programs: Singapore, London, Stockholm, Milan, Gothenburg, and most recently, New York City.
Evidence from these international cases consistently shows reductions in vehicle entries into the priced zones and measurable increases in public transit use.

\paragraph{Singapore}
Singapore was the first city to introduce an electronic congestion pricing scheme, Electronic Road Pricing, in September 1998.
ERP features time-of-day and vehicle-class differentiation, enforced through a network of gantries. At launch, vehicle entries into the CBD declined by about 15\% overall and 16\% during the morning peak \citep{lehe2019downtown}.
Studies exploiting ERP toll adjustments found immediate and sustained increases in bus ridership—roughly 12–20\% in the morning peak and about 10\% in the evening peak—especially in lower-income areas \citep{agarwal2016impact}.

\paragraph{London}
The London Congestion Charge was introduced in February 2003.
The program established a 22 km$^2$ charging zone with a daily fee for unlimited travel within the zone during daytime hours.
Enforced by automatic number plate recognition cameras, the scheme reduced car and minicab entries by roughly one-third at the start (2003 vs. 2002), and this effect persisted for several years.
Public transport use increased in parallel, with Underground exits at central stations rising about 10\% by 2007 and bus patronage remaining above pre-charge levels \citep{TfL_2008_LondonCC_SixthReport,green2016traffic}.

\paragraph{Stockholm}
Stockholm trialed a Congestion Tax in 2006 before making it permanent in 2007.
Charges varied by time of day and applied to vehicles crossing a cordon around the inner city. During the trial, cordon crossings fell sharply (about -28\% in January year-over-year), stabilizing at -20\% by May.
Since full implementation, traffic flows have remained stable despite population and economic growth, showing long-term effectiveness \citep{borjesson2012stockholm,borjesson2018long}.

\paragraph{Milan}
Milan first implemented Ecopass in 2008, an emissions-indexed access fee for the Cerchia dei Bastioni area.
In 2012, it transitioned to Area C, a simpler flat access charge that banned high-emission vehicles outright.
Ecopass reduced vehicle entries by about 14\% in its first year, growing to 17\% by 2011, but many vehicles shifted into exempt categories.
Area C achieved further reductions and increased the proportion of vehicles paying the charge (from 22\% under Ecopass to 56\% under Area C), while also improving traffic speeds \citep{gibson2015effects}.

\paragraph{Gothenburg}
Gothenburg launched its Congestion Tax in 2013 as part of a regional transport investment package.
Crossings dropped immediately but partially rebounded, stabilizing around 600,000 per day, compared to 657,000 before launch. Later toll increases produced only modest additional effects.
Travel-time savings were relatively small, reflecting Gothenburg’s lower baseline congestion \citep{borjesson2015gothenburg,lehe2019downtown}.

\paragraph{New York City}
New York became the first U.S. city to adopt congestion pricing with the Central Business District Tolling Program.
Authorized by the 2019 MTA Reform and Traffic Mobility Act, the program imposes charges for vehicles entering Manhattan at or below 60th Street.
The adopted toll schedule, recommended in 2023 and approved in 2024, began with a \$9 peak toll for cars, with planned increases through 2031 \citep{mta_congestion_2025,cook2025short}.
Early indicators suggest the policy reduced vehicle entries, improved in-zone bus speeds, and boosted transit ridership compared to 2024 \citep{MTA_2025_SixMonths_CongestionPricing}.
However, these initial findings are largely descriptive and do not establish how observed travel behavior compares with no-policy expectations.

Across international experiences, congestion pricing has consistently reduced traffic volumes and increased transit use, though the magnitude and persistence of these effects vary substantially across contexts \citep{eliasson2008lessons, gallego2013effect}.
Empirical evidence from Stockholm, London, Singapore, and Milan shows that well-enforced pricing schemes can reduce vehicle entries by 10–30\% and yield significant gains in bus speeds and reliability, while revenue reinvestment in public transport further enhances long-term mode shifts \citep{eliasson2017congestion}.
Yet the policy’s effectiveness and equity implications depend critically on local conditions, including charge structure, network redundancy, enforcement design, and the availability of high-quality transit alternatives.

New York presents a particularly complex case.
Its transit network is among the largest and most interconnected globally, and travel demand is shaped by strong transit dependency, regional commuting flows, and dense land use patterns.
Such systemwide interdependencies complicate empirical evaluation.
Unlike earlier congestion pricing implementations, New York offers no obvious control group of unaffected travelers or areas \citep{baghestani2020evaluating, morton2025impact}.
Furthermore, spillover effects across bridges, tunnels, and transit modes, coupled with pandemic-era recovery dynamics and concurrent service adjustments, make simple before-and-after comparisons misleading.

To assess post-policy changes more rigorously, it is therefore necessary to move beyond descriptive statistics toward a robust counterfactual modeling framework.
This raises a central challenge: How can we estimate what would have happened in the absence of the policy—i.e., construct a credible counterfactual—when no clearly comparable untreated group exists?
Next, we review established quasi-experimental methods in policy analysis, highlighting both their contributions and limitations in the context of large-scale urban interventions such as New York’s congestion pricing program.

\subsection{Quasi-experimental approaches in transportation policy analysis}
\label{sec:si_litreview_quasi}

Quasi-experimental research designs aim to isolate policy-related changes from correlations driven by confounding factors, selection bias, or natural variation.
In transportation and urban policy research, several econometric strategies have been widely employed to achieve this goal, including difference-in-differences (DiD), regression discontinuity design (RDD), synthetic controls, and panel fixed effects models.

DiD compares changes in outcomes between treated and untreated units before and after a policy, under the key assumption of \textit{parallel trends}.
It has been used extensively in congestion pricing evaluations by contrasting treated and control areas or time periods \citep{borjesson2012stockholm, gibson2015effects, lehe2019downtown}.
More recent work has attempted to improve DiD robustness by allowing for dynamic treatment timing, heterogeneous effects, or staggered adoption \citep{athey2022design, callaway2021difference}.
Nevertheless, its validity depends on the availability of a plausible control group and stable pre-trends—conditions often violated in complex urban systems such as New York, where pandemic recovery, service changes, and behavioral spillovers challenge parallel trend assumptions.

RDD identifies discontinuities around sharp policy thresholds, such as toll zone boundaries or eligibility cutoffs \citep{angrist2009mostly, lee2010regression}.
It is appealing for settings where assignment to treatment is discontinuous, but its credibility hinges on limited manipulation near the threshold and the absence of spillovers.
In congestion pricing contexts, where behavior may adjust systemwide—through rerouting, time-shifting, or mode switching—discontinuities become blurred, weakening the validity of RDD assumptions \citep{duranton2018urban}.

Synthetic control methods construct weighted combinations of untreated units to match the pre-intervention trajectory of the treated unit \citep{abadie2010synthetic}.
This has been particularly influential in single-unit policy interventions, such as the London or Stockholm pricing schemes.
Panel fixed effects models similarly leverage variation across units and time to control for unobserved heterogeneity \citep{wooldridge2010econometric}.
However, both approaches depend on the presence of comparable untreated units. When an entire metropolitan system is affected—as in New York—this assumption fails, and synthetic controls may produce biased estimates due to poor match quality \citep{arkhangelsky2021synthetic, doudchenko2016balancing}.

These limitations underscore a central challenge in studying New York City’s congestion pricing program: the absence of a clearly defined control group and the presence of strong spatial and temporal spillovers.
Traditional methods struggle to construct a credible counterfactual baseline in such a setting, where policy-related changes may ripple across regions, modes, and time frames.

In light of these challenges, recent work in econometrics and machine learning has explored strategies that move beyond explicit control groups.
Forecasting-based approaches—especially those incorporating probabilistic or Bayesian inference—offer a flexible framework for constructing expected no-policy demand without relying on untreated units.
By learning temporal dynamics from pre-policy data and quantifying the expected range of outcomes, these models estimate counterfactual trajectories and enable statistical testing of observed deviations \citep{brodersen2015inferring, schulam2017reliable, hill2020bayesian}.
These methods have been applied in public health, labor economics, and increasingly, in urban analysis, where systemwide interventions complicate the construction of comparison groups.

The next section reviews this class of forecasting methods, highlighting how classical and machine learning-based time series forecasting models can support uncertainty-aware counterfactual analysis in high-dimensional, spatially diffuse transportation settings.

\subsection{Forecasting-based counterfactual analysis}
\label{sec:si_litreview_counterfactual}

Forecasting-based counterfactual approaches aim to estimate what would have happened in the absence of an intervention by learning temporal dynamics from pre-policy data. This strategy is particularly relevant in urban transportation settings where finding a comparable control group is infeasible due to the scale, interdependence, and complexity of the system.

A foundational example is the Bayesian Structural Time Series (BSTS) model \citep{brodersen2015inferring}, which has been widely adopted for estimating counterfactual outcomes around events and policies using time series data. BSTS provides posterior distributions over counterfactual outcomes, enabling hypothesis testing via prediction intervals. More recent econometric work continues to refine these ideas through semi-parametric or Bayesian frameworks with improved interpretability and robustness to structural breaks \citep{hill2020bayesian, cerqua2024causal}.

In parallel, the machine learning community has developed a range of models that extend counterfactual forecasting to high-dimensional and nonlinear settings. Neural approaches have also been developed for estimating time-varying responses in sequential data, including recurrent encoder--decoder architectures with treatment histories \citep{bicaestimating} and Bayesian deep learning models with uncertainty quantification \citep{cao2023estimating}.

Other methods adapt intervention-response modeling ideas to the time series domain by incorporating temporal structure into frameworks for heterogeneous response estimation. Examples include uplift modeling for dynamic interventions \citep{alaa2018limits}, Bayesian deep learning frameworks \citep{han2024interpretable}, and Gaussian process-based counterfactual regression \citep{rothfuss2021pacoh}. These methods emphasize the importance of both modeling dynamic response trajectories and accounting for epistemic uncertainty in the prediction process.

Across these approaches, a common theme is the use of probabilistic forecasting to construct counterfactual trajectories and assess whether observed post-treatment outcomes lie within expected confidence bounds.
This framework not only accommodates the absence of untreated comparison units but also provides a principled basis for estimating uncertainty—a key requirement for robust policy analysis in volatile, high-stakes environments.

In transportation and mobility policy analysis, applications of these techniques remain relatively limited but are growing. Recent advances in time series foundation models have further expanded the potential of counterfactual forecasting by enabling large-scale pretrained models to capture complex temporal dynamics across heterogeneous datasets. In particular, models such as Chronos \citep{ansari2024chronos} and TimesFM \citep{das2023decoder} leverage transformer-based architectures and large-scale pretraining to improve forecasting robustness and generalization across diverse time series tasks. Recent research has shown that large-scale pre-trained time series foundation models can achieve highly accurate forecasting performance \citep{ahamedzero} and have demonstrated strong success across applications in public health \citep{wang2025comparative}, finance \citep{lok2025forecasting}, and electricity markets \citep{santos2026forecasting}. Building on this emerging literature, our work integrates time series foundation models into a scalable probabilistic counterfactual forecasting pipeline to examine New York City’s congestion pricing program across multiple transportation modes and spatial units, while explicitly accounting for spatiotemporal dependencies and predictive uncertainty.

% ======================================================================
% ======================================================================
\section{Data and CRZ classification details}
\label{sec:si_data}

\subsection{Data sources}

To examine multimodal travel behavior after New York City’s congestion pricing implementation, we assemble a consistent set of bus, subway, and aggregate travel demand measures.
Specifically, we use MTA bus ridership (route--hour boardings), MTA subway ridership (station--hour entries/exits and monthly origin--destination (OD) estimates), to construct station-to-station flows. These data capture complementary aspects of transit use and enable both temporal and spatial analyses.

Table~\ref{tab:si_data_sources} summarizes all data sources and their native spatial and temporal resolutions, and the following paragraphs describe each dataset in detail.
Section~\ref{descriptive_paragraph} then reports descriptive patterns before and after policy implementation.
\begin{table}[!ht]
  \centering
  \caption{Summary of data sources used in this study. All datasets are aggregated to a daily resolution and used only for the period prior to April 30, 2025.}
  \label{tab:si_data_sources}
  \begin{adjustbox}{width=\textwidth}
    \begin{tabular}{l l l l}
      \toprule
      \textbf{Dataset} & \textbf{Description} & \textbf{Spatial Resolution} & \textbf{Original Temporal Resolution} \\
      \midrule
      MTA Subway OD Estimates & Monthly OD flows from OMNY/MetroCard & Station-to-station & Monthly \\
      MTA Subway Hourly Ridership & Entries/exits at stations & Station & Hourly \\
      MTA Bus Hourly Ridership & Boardings by route and fare class & Bus route & Hourly \\
      Replica Trip Counts OD & Trip records with census tracts ID & Census tract & Weekly \\
      \bottomrule
    \end{tabular}
  \end{adjustbox}
\end{table}

\paragraph{Bus data}
The MTA’s Bus Hourly Ridership dataset reports route-level boardings by hour (available beginning in 2025).\footnote{\url{https://catalog.data.gov/dataset/mta-bus-hourly-ridership-beginning-2025}}
Each record includes a route ID, timestamp, and ridership counts disaggregated by fare class (e.g., full fare and reduced fare).
These data support high-frequency analysis of bus demand, including weekday--weekend differences, peak-period patterns, and short-run responses to congestion pricing.

\paragraph{Subway data}
The Metropolitan Transportation Authority (MTA) publishes two complementary datasets that capture subway usage at different spatial and temporal scales.
First, the \textit{Subway Hourly Ridership dataset} reports station-level entries and exits by hour (2020--2025) and is disaggregated by fare class (e.g., OMNY and reduced fare).\footnote{\url{https://data.ny.gov/Transportation/MTA-Subway-Hourly-Ridership-2020-2024/wujg-7c2s/about_data}}

Second, the \textit{Subway Origin--Destination Ridership Estimates} provide monthly OD flows between station complexes, stratified by hour of day and day of week, and derived from OMNY and MetroCard transactions.\footnote{\url{https://www.mta.info/article/introducing-subway-origin-destination-ridership-dataset}}
We use these OD estimates to characterize spatial shifts in travel patterns and construct a daily OD series by applying the reported day-of-week profiles.

\paragraph{Trip count data}
The Replica mobility dataset provides origin--destination trip estimates across multiple transportation modes and spatial units, capturing aggregate travel patterns within the study area. \footnote{\url{https://www.replicahq.com/}} The data include trip counts by origin, destination, mode, and time period, allowing us to examine changes in overall travel demand and modal redistribution following congestion pricing. We use these records to construct OD-level mobility flows and analyze temporal and spatial changes in travel demand before and after policy implementation.

Note that all counterfactual modeling and subsequent analyses are conducted using data prior to April 30, 2025. We restrict the study period to the short term for two primary reasons. First, post-policy changes may attenuate over time as behavioral responses stabilize and travelers adapt to the policy. Second, forecasting uncertainty generally increases with the prediction horizon, which can compromise the validity of long-term counterfactual estimates.

Given these considerations, we focus on the immediate responses to congestion pricing, which are more directly measurable and particularly relevant for examining early-stage demand changes. By integrating bus, subway, and aggregate trip-count datasets, we examine post-policy changes from multiple perspectives, including public transit ridership at different spatial and temporal resolutions and broader changes in overall travel demand.

\subsection{CRZ classification}

To examine spatial heterogeneity in post-policy demand changes, we classify each transport unit by its geographic relationship to the Congestion Relief Zone (CRZ).
For spatial units that can physically span the CRZ boundary—such as bus routes and certain origin–destination (OD) connections—we define three categories: \textit{fully inside}, \textit{partially inside}, and \textit{outside}.
Units fully inside the CRZ are directly exposed to the tolling policy and associated in-zone changes (e.g., reduced traffic and improved surface speeds), whereas units fully outside the CRZ are not tolled but may respond indirectly through systemwide spillovers.
The \textit{partially inside} category captures services and trips that explicitly connect CRZ and non-CRZ areas, reflecting mixed exposure along their spatial extent.

For point-based units that do not cross the CRZ boundary—such as subway stations and census tract-level aggregates—classification is restricted to \textit{inside} versus \textit{outside} the CRZ.
This restriction ensures that spatial categories are consistent with the underlying geometry of each dataset and avoids imposing artificial boundary-crossing interpretations where they are not meaningful.

This classification scheme is deliberately intuitive, yet central to interpreting the results that follow.
Because the CRZ sits at the core of New York City’s multimodal network, congestion pricing can affect travel demand both locally within the tolling area and indirectly through diversion, mode substitution, and network-wide spillovers.
Separating \textit{inside} from \textit{outside} isolates direct policy exposure from broader system responses, while the \textit{partially inside} category highlights corridors and OD connections where responses may differ in sign, magnitude, or timing.
These distinctions help explain why some changes concentrate in and around the CRZ, while others emerge along feeder corridors and in outer-borough travel markets.

\begin{figure}[!ht]
  \centering
  \includegraphics[width=\textwidth]{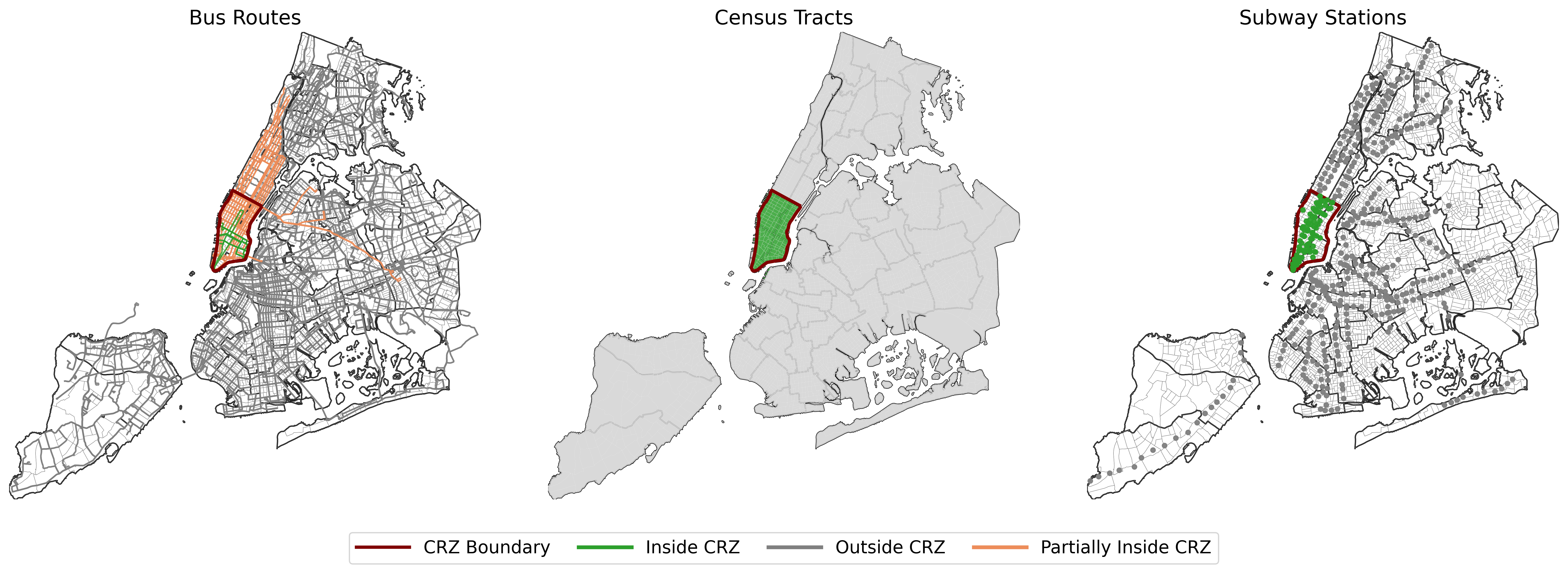}
  \caption{Geographic distribution of transport units by CRZ classification.}
  \label{fig:si_crz}
\end{figure}

Figure~\ref{fig:si_crz} visualizes the classification for bus routes, census tracts, and subway stations, and Table~\ref{tab:si_crz_counts} reports the corresponding counts.

\begin{table}[!ht]
  \centering
  \caption{Spatial distribution of each dataset by location category relative to the CRZ.}
  \label{tab:si_crz_counts}
  \begin{tabular}{l c c c}
    \toprule
    \textbf{Dataset} & \textbf{Fully inside} & \textbf{Partially inside} & \textbf{Outside} \\
    \midrule
    Bus route       & 5   & 49  & 192 \\
    Subway station  & 62  & --  & 310 \\
    Census tract    & 135 & --  & 2{,}190 \\
    \bottomrule
  \end{tabular}
\end{table}

\section{Descriptive and exploratory comparisons}
\label{descriptive_paragraph}

This section summarizes descriptive evidence on how travel demand changed after congestion pricing.
We first compare daily and day-of-week patterns across bus and subway systems, then provide additional exploratory views of spatial and temporal heterogeneity.
These comparisons motivate the counterfactual analysis that follows, but should not be interpreted as policy-attributable estimates on their own.

\subsection{Daily travel demand change}

\begin{figure}[!htbp]
  \centering
  \captionsetup{justification=centering}
  \captionsetup[subfigure]{justification=centering}

  \begin{subfigure}[t]{0.45\textwidth}
    \centering
    \includegraphics[width=\linewidth]{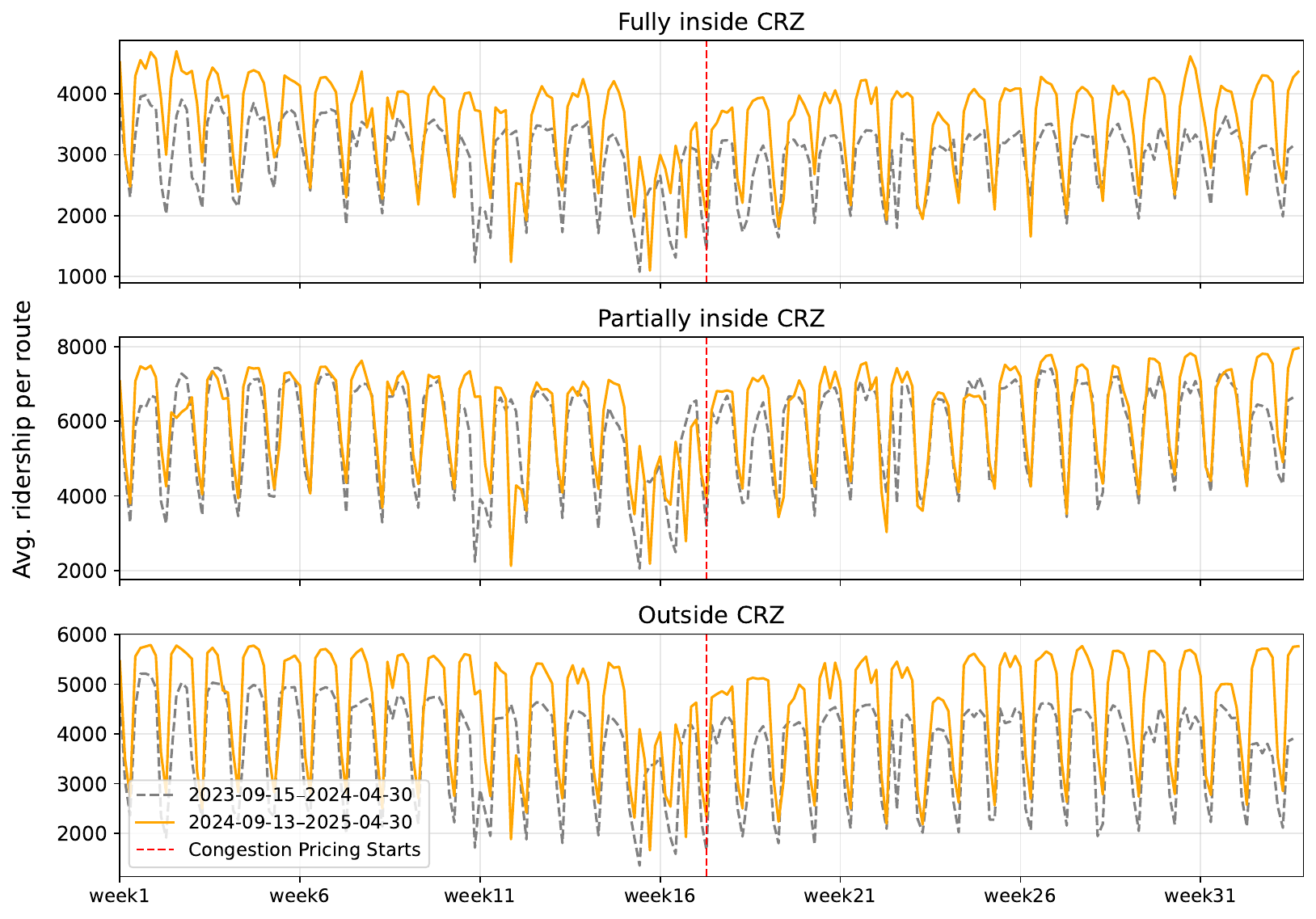}
    \subcaption{Bus: average daily ridership by CRZ classification}
    \label{fig:bycrz-bus}
  \end{subfigure}
  \hspace{0.02\textwidth}
  \begin{subfigure}[t]{0.45\textwidth}
    \centering
    \includegraphics[width=\linewidth]{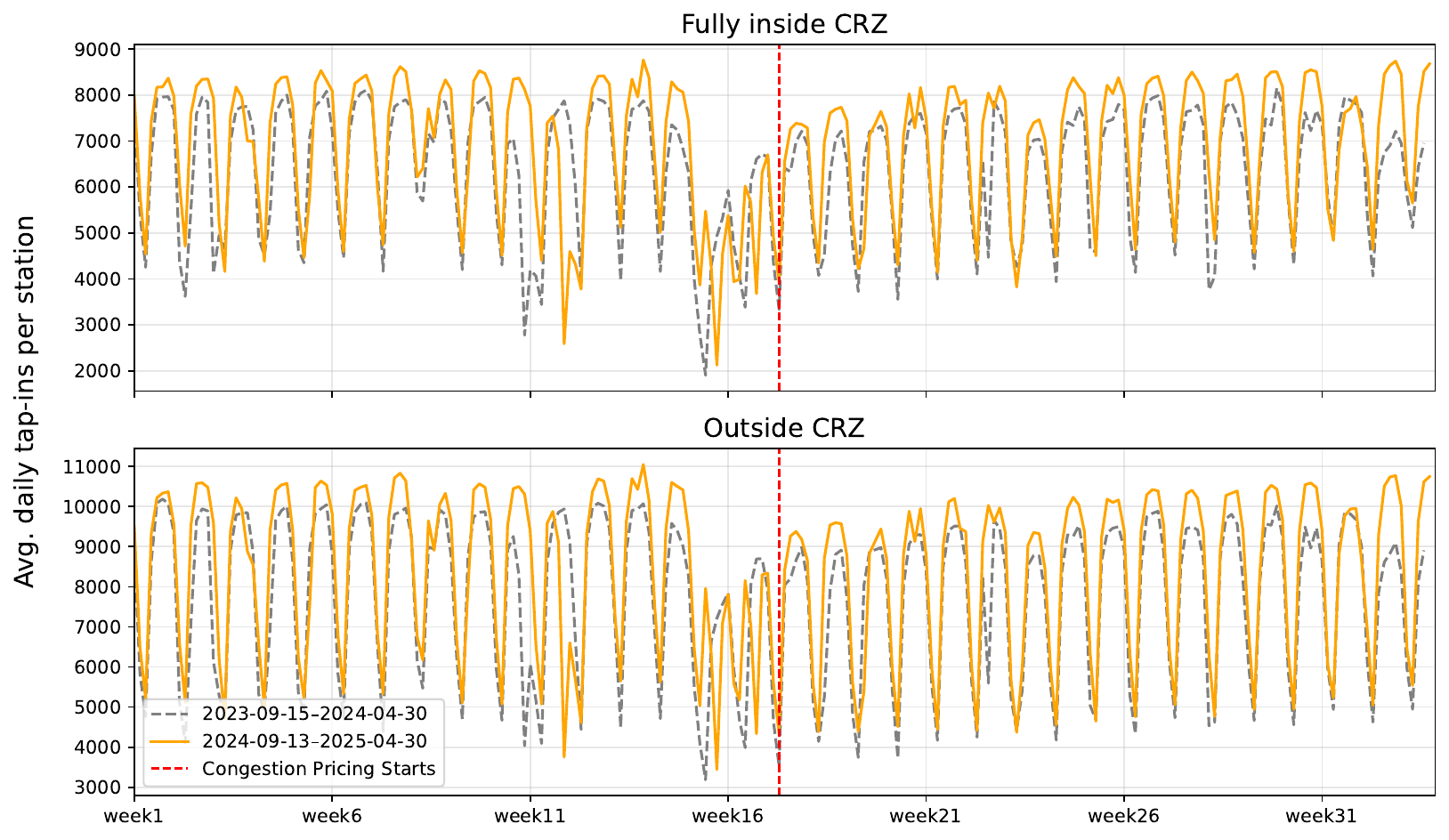}
    \subcaption{Subway: average daily tap-ins by CRZ classification}
    \label{fig:bycrz-subway}
  \end{subfigure}

  \caption{Daily demand trajectories by CRZ classification across bus and subway, comparing the post-policy period after January~5,~2025 with a seasonally matched baseline year.}
  \label{fig:crz-daily-allmodes}
\end{figure}

We assess daily travel demand responses to congestion pricing across two major modes: bus and subway. Figure~\ref{fig:crz-daily-allmodes} compares ridership trajectories before and after program implementation (January 5, 2025), disaggregated by CRZ classification: \textit{fully inside}, \textit{partially inside}, and \textit{outside} the toll zone.

\paragraph{Bus ridership}
Figure~\ref{fig:bycrz-bus} reports average daily bus ridership by CRZ category across the treatment (2024--2025, orange) and control (2023--2024, gray) periods.
Following program launch, bus use increases significantly, with the strongest gains observed on routes fully inside the CRZ (+18--19\%) and outside the CRZ (+22\%).
Routes classified as partially inside the CRZ exhibit more modest growth (approximately 6\%).
This pattern suggests that in-zone improvements in travel speed and reliability may have enhanced bus attractiveness, while ridership growth outside the CRZ is consistent with spillovers from travelers avoiding tolls or shifting modes.
The comparatively smaller increase on partially inside routes is consistent with mixed exposure to the policy and uneven benefits along these corridors.

\paragraph{Subway tap-ins}
Figure~\ref{fig:bycrz-subway} shows subway tap-in volumes over the same time windows.
While the magnitude of the increase is smaller than for buses, a clear upward shift is observed after implementation, corresponding to a systemwide rise of approximately 9\%.
Gains are broadly distributed across the network rather than concentrated in a single area, indicating that subway demand responded positively to reduced car traffic and improved multimodal conditions associated with congestion pricing.

Overall, we observe consistent increases in public transit demand following the introduction of congestion pricing.
Bus ridership exhibits the strongest and most spatially differentiated response, reflecting both direct improvements within the CRZ and heterogeneous exposure across routes.
Subway tap-ins also increase systemwide, consistent with network-wide adjustments and behavioral spillovers rather than purely localized changes.
These results highlight congestion pricing's role in reallocating travel demand toward more sustainable modes and underscore the importance of multimodal integration in policy design.

\subsection{Day-of-week travel demand change}

\begin{figure}[!htbp]
  \centering

  % ========= (a) Bus row =========
  \begin{subfigure}[t]{\textwidth}
    \centering
    \begin{subfigure}[t]{0.32\textwidth}
      \includegraphics[width=\linewidth]{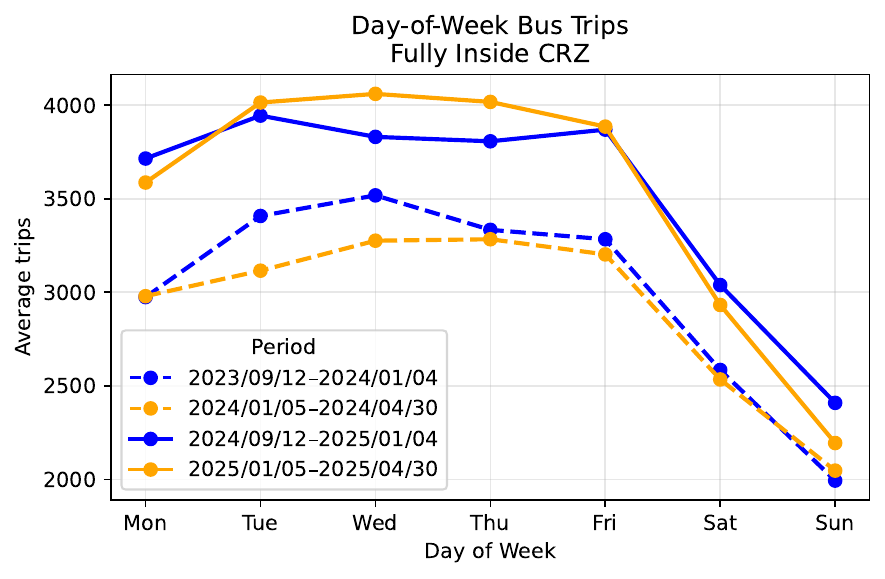}
    \end{subfigure}\hfill
    \begin{subfigure}[t]{0.32\textwidth}
      \includegraphics[width=\linewidth]{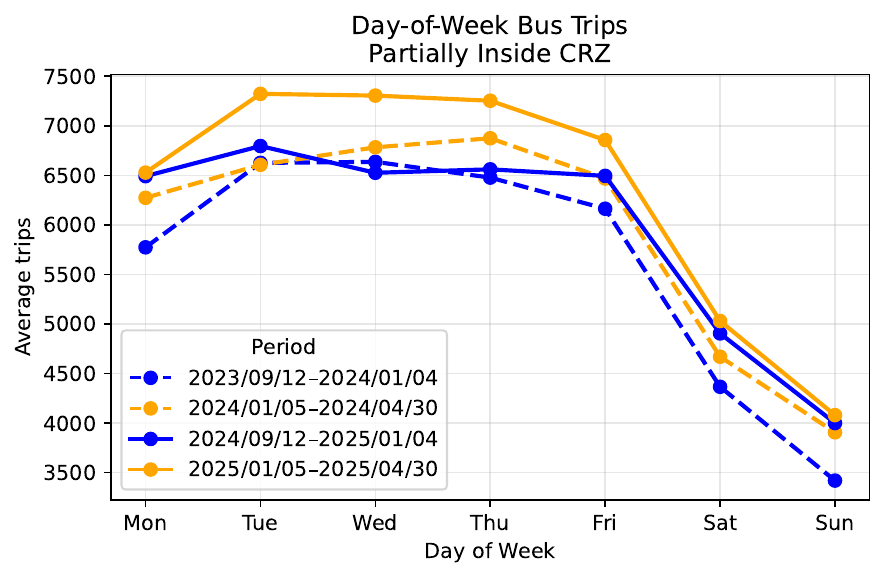}
    \end{subfigure}\hfill
    \begin{subfigure}[t]{0.32\textwidth}
      \includegraphics[width=\linewidth]{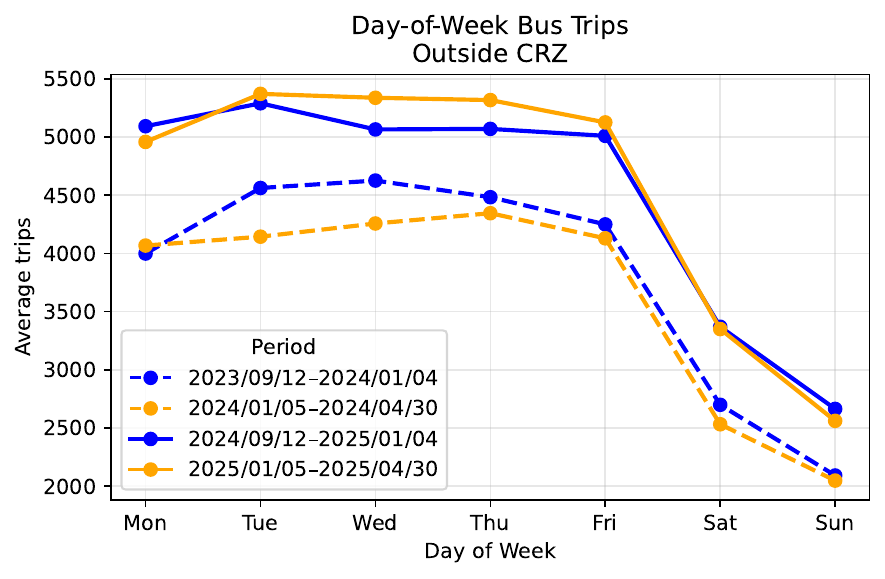}
    \end{subfigure}
    \caption{Bus ridership patterns by CRZ division}
    \label{fig:dow_bus_crz}
  \end{subfigure}

  \vspace{1em}

  % ========= (b) Subway row =========
  \begin{subfigure}[t]{\textwidth}
    \centering
    \begin{subfigure}[t]{0.32\textwidth}
      \includegraphics[width=\linewidth]{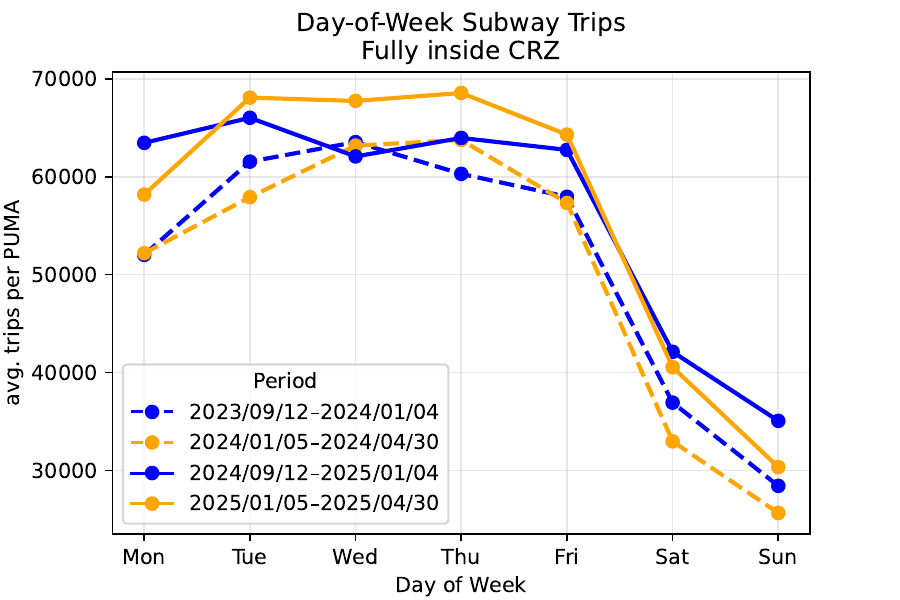}
    \end{subfigure}
    % \begin{subfigure}[t]{0.32\textwidth}
    %   \includegraphics[width=\linewidth]{figures/final_subway/dow_subway_02_partial_CRZ_avg.pdf}
    % \end{subfigure}\hfill
    \begin{subfigure}[t]{0.32\textwidth}
      \includegraphics[width=\linewidth]{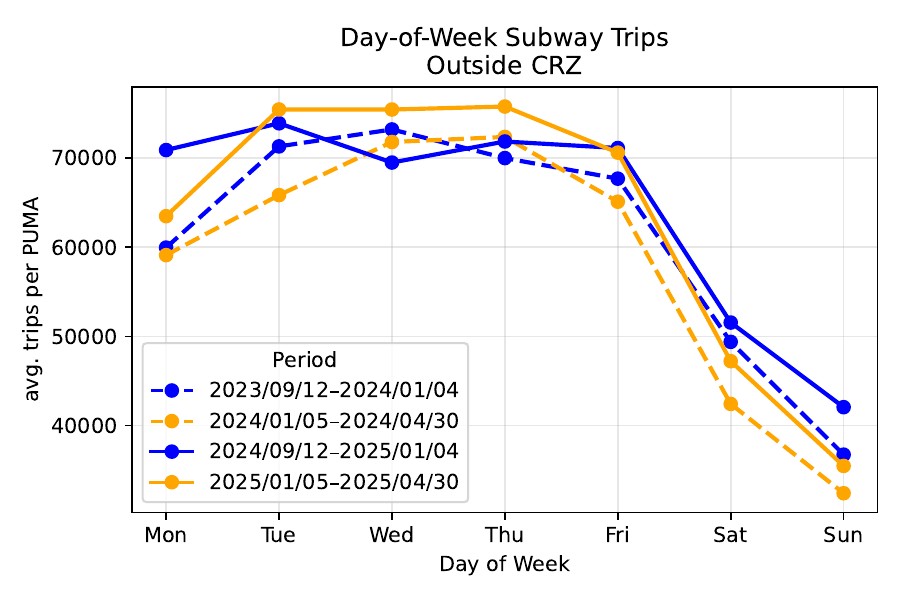}
    \end{subfigure}
    \caption{Subway ridership patterns by CRZ division}
    \label{fig:dow_subway_crz}
  \end{subfigure}

  \vspace{1em}

  \caption{Average day-of-week ridership patterns before and after congestion pricing, stratified by CRZ division, for (a) buses and (b) subways based on corresponding spatial resolutions. Curves compare pre-policy and post-policy periods using seasonally matched baseline years.}
  \label{fig:dow_bus_subway}
\end{figure}

We next analyze how congestion pricing affected public transit demand across days of the week, disaggregated by mode and CRZ classification.
Figure~\ref{fig:dow_bus_subway} presents average weekday and weekend ridership profiles before and after the policy's implementation on January~5,~2025, using both 2024--2025 observations and seasonally matched 2023--2024 baseline years.
Orange indicates post-policy outcomes, while blue denotes the baseline.

\paragraph{Bus ridership}
As shown in Figure~\ref{fig:dow_bus_crz}, bus ridership increased substantially on weekdays across all spatial groups.
Average weekday ridership rose by 20.8\% on routes fully inside the CRZ, by 6.7\% on routes partially inside the CRZ, and by 25.5\% on routes outside the CRZ.
Weekend ridership remained comparatively low and stable across all groups.
These patterns indicate a strongly weekday-dominant response, suggesting that congestion pricing improved weekday bus reliability and attractiveness both within and beyond the toll zone.
The pronounced response outside the CRZ is consistent with spillovers from travelers avoiding tolls or shifting modes, while the more modest gains on partially inside routes likely reflect mixed exposure and uneven benefits along those corridors.

\paragraph{Subway tap-ins}
Figure~\ref{fig:dow_subway_crz} shows subway tap-in profiles characterized by a clear weekday--weekend separation.
Weekday tap-ins increased by 12.7\% within the CRZ and by 8.4\% outside the CRZ, while weekend volumes remained largely unchanged.
The strongest gains occur from Tuesday through Thursday, aligning with peak commuting days.
These results indicate that subway ridership increased primarily on weekdays, with changes concentrated in areas either fully exposed to the toll zone or outside it.

Across both modes, weekday demand increased markedly after the introduction of congestion pricing, while weekend activity exhibited limited change.
Bus ridership shows the most spatially heterogeneous response, reflecting differential exposure across routes.
Subway responses are similarly dominated by weekday commuting behavior, with increases observed both inside and outside the CRZ.
Taken together, these findings support the interpretation that congestion pricing primarily reshaped regular workweek travel patterns rather than discretionary travel, inducing meaningful modal shifts during core commuting days.

\subsection{Additional descriptive and exploratory analyses}
\label{sec:appendix_descriptive}

We next provide complementary descriptive analyses that support the main counterfactual forecasting results.
These figures summarize spatial and temporal ridership patterns across modes, routes, and time dimensions, highlighting heterogeneity that motivates and contextualizes the probabilistic counterfactual analysis.
As above, these results are descriptive and should be read as exploratory evidence rather than policy-attributable estimates.

% ============================================================
\subsubsection{Mode-specific descriptive patterns}
\label{sec:appendix_mode}

\begin{figure}[H]
\centering
\includegraphics[width=0.5\textwidth]{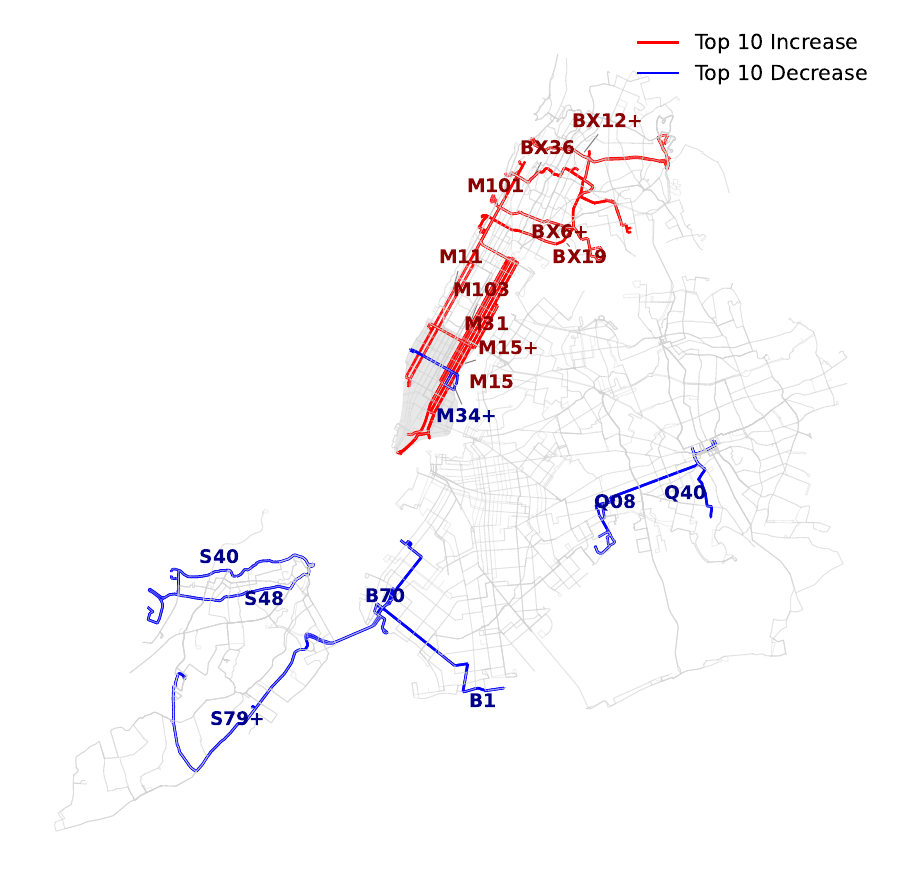}
\caption{Spatial distribution of the top 10 bus routes with the largest ridership increases and decreases.}
\label{fig:bus_top10}
\end{figure}
\FloatBarrier

Figure~\ref{fig:bus_top10} maps the routes with the most pronounced ridership changes.
Red segments denote increases and blue segments denote decreases.
The spatial clustering of increases along core Manhattan and Bronx corridors
is consistent with systemwide patterns observed in the main analysis,
while localized declines suggest more limited adjustments on selected outer-borough routes.

% ============================================================
\subsubsection{Spatial and temporal heterogeneity}
\label{sec:appendix_spatial}

\begin{figure}[H]
  \centering

  \newlength{\RowAHeight}\setlength{\RowAHeight}{0.34\textwidth}
  \newlength{\RowBHeight}\setlength{\RowBHeight}{0.17\textwidth}

  \begin{subfigure}[t]{0.49\textwidth}
    \centering
    \makebox[\linewidth][c]{%
      \includegraphics[height=\RowAHeight,keepaspectratio]{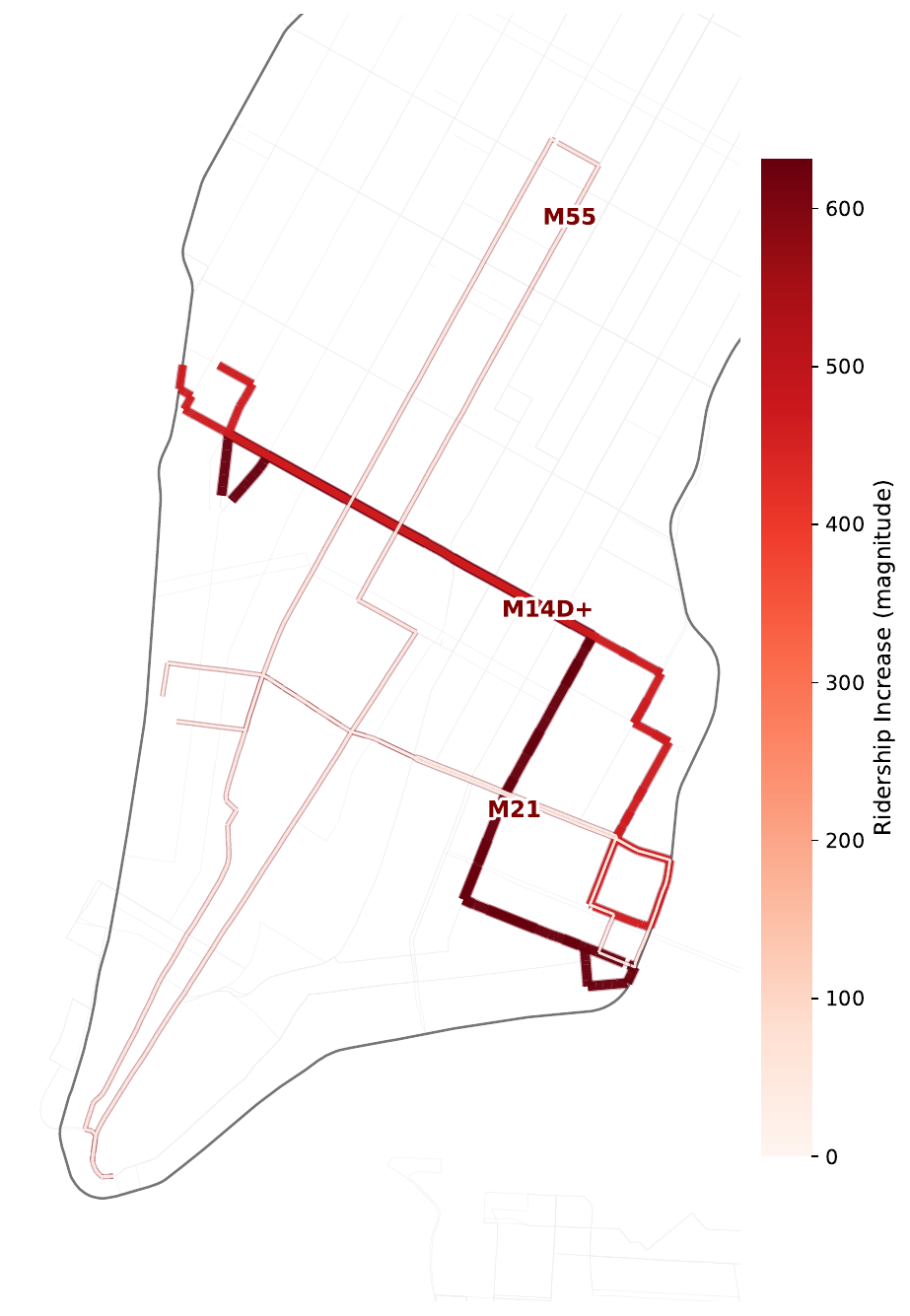}}
    \subcaption{Fully-inside CRZ routes: weekday ridership increase, 01:00--08:00}
    \label{fig:bus-map-inc}
  \end{subfigure}\hfill
  \begin{subfigure}[t]{0.49\textwidth}
    \centering
    \makebox[\linewidth][c]{%
      \includegraphics[height=\RowAHeight,keepaspectratio]{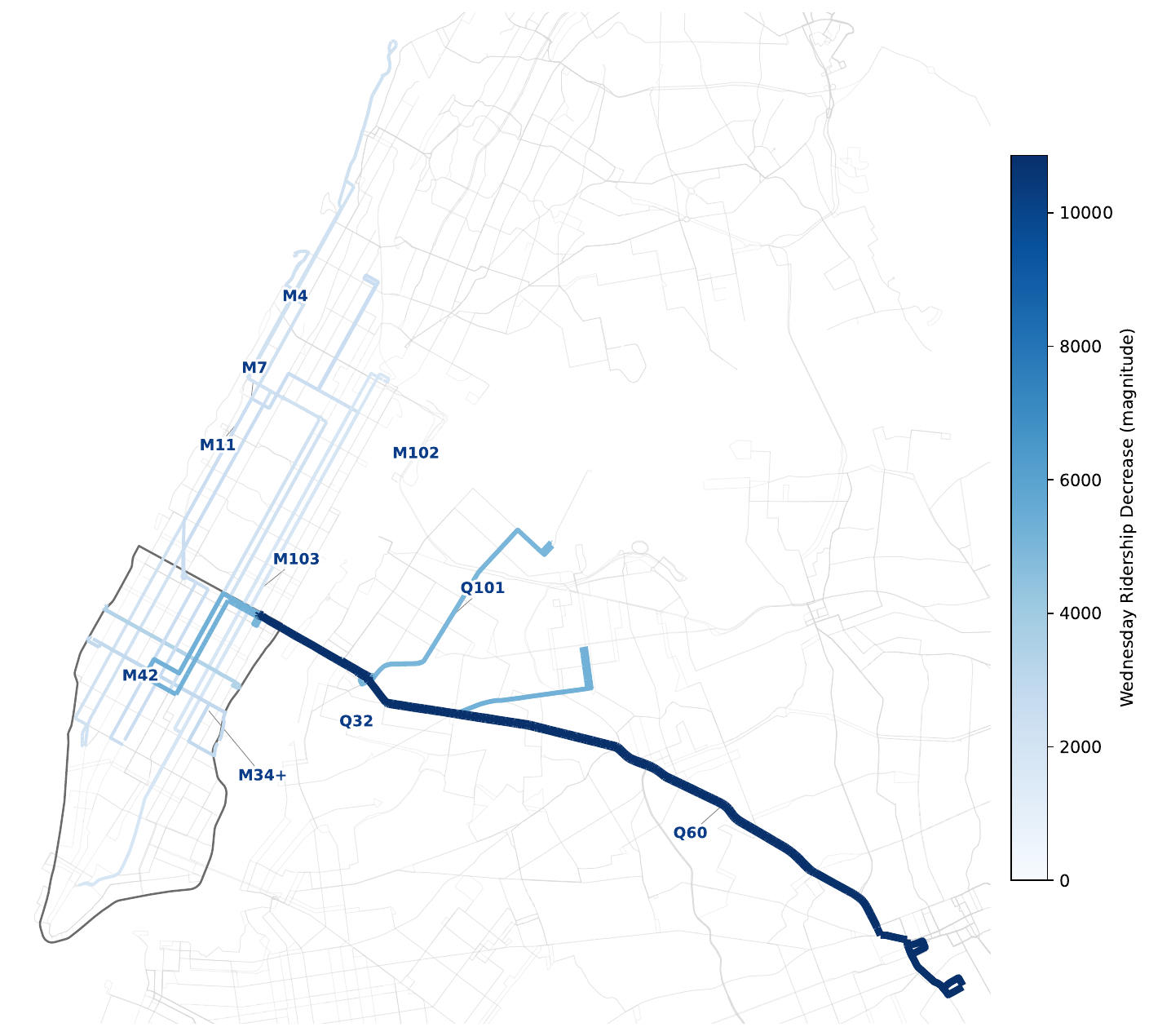}}
    \subcaption{Partially-inside CRZ routes: weekday ridership decrease, 23:00--06:00}
    \label{fig:bus-map-dec}
  \end{subfigure}

  \vspace{0.9em}

  \begin{subfigure}[t]{0.5\textwidth}
    \centering
    \makebox[\linewidth][c]{%
      \includegraphics[height=\RowBHeight,keepaspectratio]{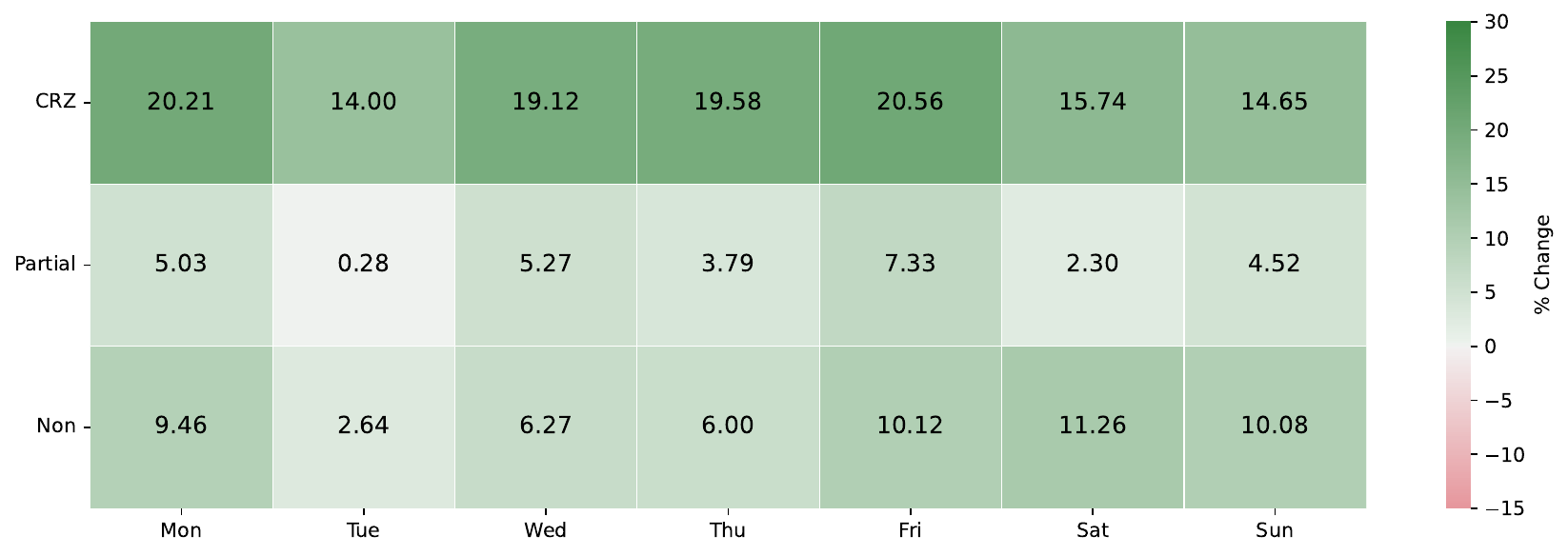}}
    \subcaption{Day-of-week change (\%)}
    \label{fig:bus-dow}
  \end{subfigure}
  \hfill
  \begin{subfigure}[t]{0.5\textwidth}
    \centering
    \makebox[\linewidth][c]{%
      \includegraphics[height=\RowBHeight,keepaspectratio]{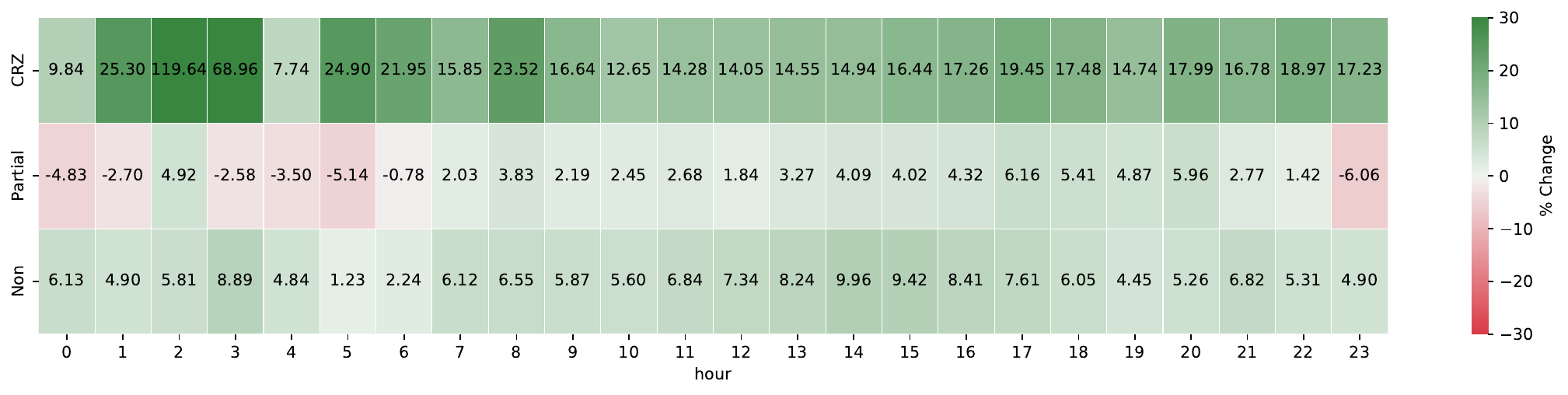}}
    \subcaption{Hourly change (\%)}
    \label{fig:bus-hourly}
  \end{subfigure}

  \caption{Spatial and temporal heterogeneity in bus ridership changes after congestion pricing.}
  \label{fig:bus-4grid}
\end{figure}

Figure~\ref{fig:bus-4grid} summarizes bus ridership changes across both space and time.
Early-morning increases are most pronounced on routes fully inside the CRZ,
particularly along core east--west corridors.
Overnight decreases are concentrated on routes that straddle the CRZ boundary.
Aggregated patterns reveal stronger weekday than weekend growth
and a pronounced increase during the morning peak,
while off-peak effects remain comparatively modest.

% ============================================================
\subsubsection{Temporal patterns by hour and day}
\label{sec:appendix_temporal}

\begin{figure}[H]
\centering

\begin{subfigure}[t]{\textwidth}
  \centering
  \includegraphics[width=\textwidth]{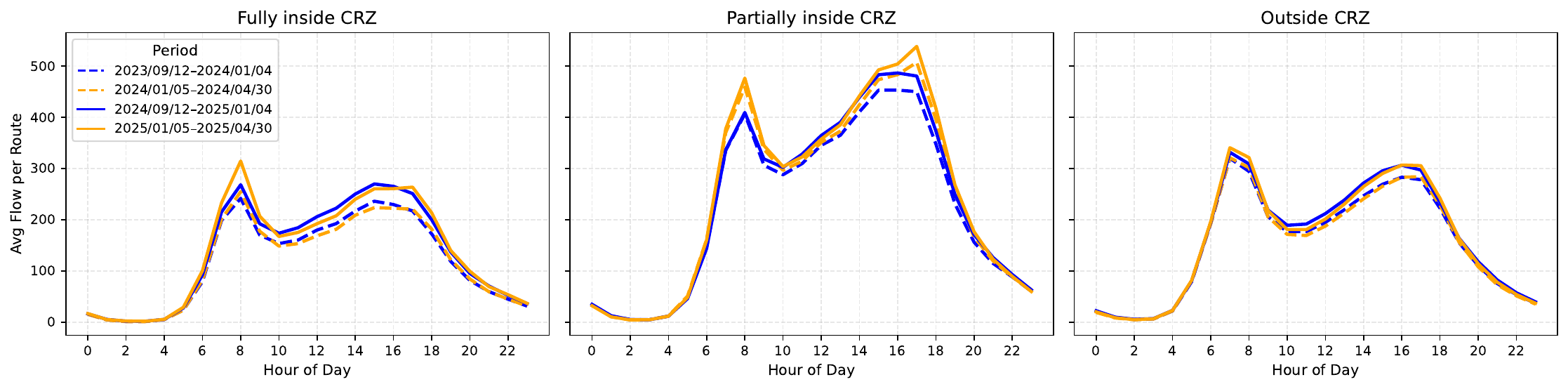}
  \caption{Bus ridership by hour of day and CRZ classification}
  \label{fig:hourly-crz-bus}
\end{subfigure}

\vspace{1em}

\begin{subfigure}[t]{\textwidth}
  \centering
  \includegraphics[width=\textwidth]{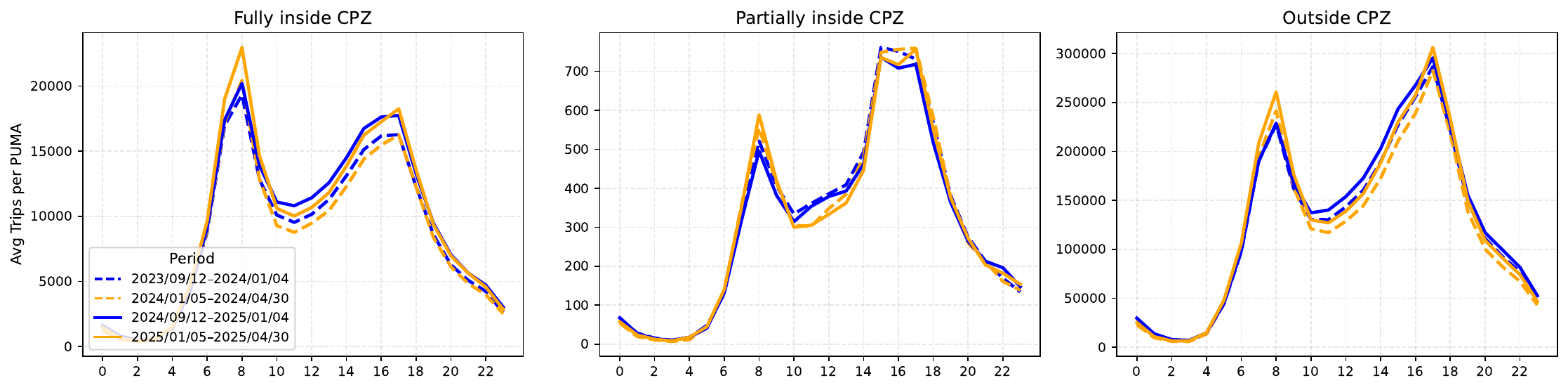}
  \caption{Subway ridership by hour of day and CRZ classification}
  \label{fig:hourly-crz-subway}
\end{subfigure}

\caption{Hourly ridership patterns by CRZ classification for buses and subways.}
\label{fig:hourly-crz-combined}
\end{figure}

Figure~\ref{fig:hourly-crz-combined} compares hourly ridership profiles before and after congestion pricing.
Both modes exhibit the strongest post-policy increases during weekday peak hours,
particularly within and near the CRZ,
while changes during off-peak hours are smaller.

\section{Problem Notation}

Table~\ref{tab:notation} summarizes the notation used throughout the study. We use $i$ to index spatial units, such as bus routes, subway stations, or census tracts, and $t$ to index time. The notation distinguishes observed travel demand, raw TimesFM forecasts, and calibrated counterfactual outcomes under the no-policy scenario. It also defines the residual-based calibration quantities, prediction intervals, pointwise and cumulative deviations from expected demand, relative changes, and significance measures used in the subsequent counterfactual analysis.

\begin{sidewaystable}[!htbp]
  \centering
  \caption{Notation Summary}
  \label{tab:notation}
  \scriptsize
  \begin{tabular}{ll|ll}
    \toprule
    Symbol & Definition & Symbol & Definition \\
    \midrule
    $i$ & Spatial unit index, such as bus route, subway station, or census tract 
    & $t$ & Time index \\

    $t_0$ & Congestion pricing implementation date 
    & $T_{\text{post}}$ & End of the post-policy evaluation period \\

    $T$ & Generic post-policy time point, $t_0 \leq T \leq T_{\text{post}}$ 
    & $N$ & Total number of spatial units \\

    $N_t$ & Number of observed units at time $t$ 
    & $M$ & Total number of validation observations across units and time \\

    $y_{i,t}$ & Observed travel demand for unit $i$ at time $t$ 
    & $\mu_{i,t}$ & Raw TimesFM forecast for unit $i$ at time $t$ \\

    $\hat{y}^{c}_{i,t}$ & Calibrated counterfactual demand under the no-policy scenario 
    & $\hat{y}^{L,\alpha}_{i,t}$ & Lower bound of the calibrated $\alpha$-level prediction interval \\

    $\hat{y}^{U,\alpha}_{i,t}$ & Upper bound of the calibrated $\alpha$-level prediction interval 
    & $\alpha$ & Nominal prediction interval coverage level \\

    $z_{\alpha}$ & Two-sided Gaussian critical value for coverage level $\alpha$ 
    & $f_{\theta}$ & Base time series foundation model forecaster \\

    $\mathbf{y}_{i,1:t_0-1}$ & Pre-policy demand history for unit $i$ 
    & $r_{i,t}$ & Validation residual, defined as $y_{i,t} - \mu_{i,t}$ \\

    $\mathcal{V}_i$ & Validation observations available for unit $i$ 
    & $b_i$ & Unit-specific residual intercept estimated from validation residuals \\

    $\tilde{r}_{i,t}$ & De-biased validation residual after removing $b_i$ 
    & $\mathbf{x}_{i,t}$ & Calibration feature vector for unit $i$ at time $t$ \\

    $Q_q(\cdot)$ & Conditional quantile function at quantile level $q$ 
    & $\boldsymbol{\beta}_q$ & Quantile regression coefficient vector for quantile level $q$ \\

    $\hat{d}^{q}_{i,t}$ & Predicted residual correction at quantile level $q$ 
    & $\tau_{i,t}$ & Pointwise deviation from expected demand, $y_{i,t} - \hat{y}^{c}_{i,t}$ \\

    $\tau^{L}_{i,t}$ & Lower bound of the pointwise deviation 
    & $\tau^{U}_{i,t}$ & Upper bound of the pointwise deviation \\

    $\mathbbm{1}^{\text{sig}}_{i,t}$ & Indicator for statistically significant unit-time effect 
    & $\hat{\sigma}_{i,t}$ & Implied standard deviation recovered from the prediction interval \\

    $\Tau_{i,T}$ & Cumulative deviation from expected demand for unit $i$ up to time $T$ 
    & $\Tau^{L}_{i,T}$ & Lower bound of the cumulative deviation \\

    $\Tau^{U}_{i,T}$ & Upper bound of the cumulative deviation 
    & $\rho_{i,T}$ & Cumulative relative change relative to the counterfactual baseline \\

    $\rho^{L}_{i,T}$ & Lower bound of the cumulative relative change 
    & $\rho^{U}_{i,T}$ & Upper bound of the cumulative relative change \\

    $\pi^{\text{sig}}_i$ & Share of significant post-policy observations for unit $i$ 
    & $\pi^{\text{excess}}_i$ & Significance share beyond the nominal false-positive rate \\

    $\bar{\tau}_i$ & Average daily deviation for unit $i$ 
    & $\bar{\tau}_t$ & Cross-unit average deviation at time $t$ \\

    $\bar{\tau}$ & Average deviation across all unit-time observations 
    & $\bar{\Tau}_T$ & Cumulative cross-unit average deviation up to time $T$ \\

    \bottomrule
  \end{tabular}
\end{sidewaystable}

% ======================================================================

\section{Core Methods}
\label{sec:si_methods}

\subsection{Problem definition}
\label{sec:si_problem_definition}

Let $y_{i,t}$ denote the observed travel demand for spatial unit $i$ at time $t$, where $i \in \{1,\dots,N\}$ indexes bus routes, subway stations, or census tracts depending on the dataset. Let $t_0$ denote the implementation date of congestion pricing. The objective is to estimate the counterfactual demand trajectory that would have been observed after $t_0$ in the absence of the policy:
\begin{equation}
  \hat{y}^{c}_{i,t},
  \quad t \geq t_0,
\end{equation}
together with a calibrated prediction interval
\begin{equation}
  \left[
  \hat{y}^{L,\alpha}_{i,t},
  \hat{y}^{U,\alpha}_{i,t}
  \right],
\end{equation}
where $\alpha$ denotes the nominal coverage level, such as $\alpha=0.90$.

For each spatial unit $i$, we use only pre-policy observations
\begin{equation}
  \mathbf{y}_{i,1:t_0-1}
  =
  \{y_{i,1},y_{i,2},\dots,y_{i,t_0-1}\}
\end{equation}
to construct a no-policy baseline. The base forecaster $f_{\theta}$ produces an initial point forecast:
\begin{equation}
  \mu_{i,t}
  =
  f_{\theta}
  \left(
  \mathbf{y}_{i,1:t_0-1}
  \right),
  \quad t \geq t_0.
\end{equation}

The pointwise deviation from expected demand is defined as:
\begin{equation}
  \tau_{i,t}
  =
  y_{i,t} - \hat{y}^{c}_{i,t}.
  \label{eq:si_tau}
\end{equation}
A positive value indicates that observed demand exceeds the no-policy counterfactual, while a negative value indicates that observed demand is below the predicted baseline.

\subsection{Hierarchical quantile calibration}
\label{sec:si_hqc}

We split the pre-policy period into a context window and a validation window. The base time series foundation model is used to generate forecasts for the validation period. For each unit $i$ and validation time $t$, we compute:
\begin{equation}
  r_{i,t}
  =
  y_{i,t} - \mu_{i,t}.
  \label{eq:si_residual}
\end{equation}

The key forecasting assumption is that the conditional structure of forecast errors learned from the validation period remains informative for the post-policy no-policy counterfactual. Both unit-specific intercepts and pooled quantile regressions are estimated using only validation residuals from the pre-policy period. This assumption does not rule out all contemporaneous shocks; rather, it defines the conditions under which deviations from calibrated expected demand can be interpreted as unusual post-policy changes.

We first estimate a unit-specific residual intercept:
\begin{equation}
  b_i
  =
  \frac{1}{|\mathcal{V}_i|}
  \sum_{t \in \mathcal{V}_i}
  r_{i,t},
  \label{eq:si_intercept}
\end{equation}
where $\mathcal{V}_i$ denotes the validation observations available for unit $i$. The de-biased residual is:
\begin{equation}
  \tilde{r}_{i,t}
  =
  r_{i,t} - b_i.
  \label{eq:si_debiased_residual}
\end{equation}

We then fit pooled quantile regression models:
\begin{equation}
  Q_q
  \left(
  \tilde{r}_{i,t}
  \mid
  \mathbf{x}_{i,t}
  \right)
  =
  \mathbf{x}_{i,t}^{\top}\boldsymbol{\beta}_q,
  \quad q \in \{0.05,0.50,0.95\},
  \label{eq:si_qr}
\end{equation}
where $\mathbf{x}_{i,t}$ is a vector of calibration features. In our implementation, $\mathbf{x}_{i,t}$ includes day-of-week indicators, month indicators, federal holiday indicators, the raw model forecast $\mu_{i,t}$, the log historical demand level of each unit, and interactions between day-of-week indicators and log demand level.

For a post-policy observation, the calibrated residual quantile is:
\begin{equation}
  \hat{d}^{q}_{i,t}
  =
  b_i
  +
  \mathbf{x}_{i,t}^{\top}
  \hat{\boldsymbol{\beta}}_q.
  \label{eq:si_delta}
\end{equation}
The calibrated point forecast is:
\begin{equation}
  \hat{y}^{c}_{i,t}
  =
  \mu_{i,t}
  +
  \hat{d}^{0.50}_{i,t}.
  \label{eq:si_calibrated_mean}
\end{equation}
The calibrated prediction interval is:
\begin{equation}
  \left[
  \hat{y}^{L,\alpha}_{i,t},
  \hat{y}^{U,\alpha}_{i,t}
  \right]
  =
  \left[
  \mu_{i,t} + \hat{d}^{0.05}_{i,t},
  \mu_{i,t} + \hat{d}^{0.95}_{i,t}
  \right],
  \label{eq:si_calibrated_interval}
\end{equation}
for a nominal 90\% interval.

\subsection{Evaluation metrics}
\label{sec:si_evaluation_metrics}

We evaluate forecasting performance on a held-out pre-policy validation window. Let $M$ denote the total number of validation observations across units and time. We assess point prediction accuracy using RMSE, MAE, and SMAPE:
\begin{align}
  \mathrm{RMSE}
  &=
  \sqrt{
  \frac{1}{M}
  \sum_{m=1}^{M}
  (y_m-\hat{y}_m)^2
  },
  \label{eq:si_rmse} \\
  \mathrm{MAE}
  &=
  \frac{1}{M}
  \sum_{m=1}^{M}
  |y_m-\hat{y}_m|,
  \label{eq:si_mae} \\
  \mathrm{SMAPE}
  &=
  \frac{100\%}{M}
  \sum_{m=1}^{M}
  \frac{
  |y_m-\hat{y}_m|
  }{
  (|y_m|+|\hat{y}_m|)/2
  }.
  \label{eq:si_smape}
\end{align}

We evaluate uncertainty calibration using the empirical coverage rate:
\begin{equation}
  \mathrm{ECR}_{\alpha}
  =
  \frac{1}{M}
  \sum_{m=1}^{M}
  \mathbbm{1}
  \left(
  y_m
  \in
  \left[
  \hat{y}^{L,\alpha}_{m},
  \hat{y}^{U,\alpha}_{m}
  \right]
  \right),
  \label{eq:si_ecr}
\end{equation}
where $\mathbbm{1}(\cdot)$ is an indicator function. A well-calibrated model should satisfy $\mathrm{ECR}_{\alpha}\approx \alpha$.

\subsection{Pointwise deviation and significance}
\label{sec:si_pointwise_effect}

For each unit $i$ and post-policy time $t$, the pointwise deviation from expected demand is:
\begin{equation}
  \tau_{i,t}
  =
  y_{i,t}
  -
  \hat{y}^{c}_{i,t}.
  \label{eq:si_point_effect}
\end{equation}
The uncertainty bounds for the pointwise deviation are obtained by comparing the observed outcome with the counterfactual prediction interval:
\begin{align}
  \tau^{L}_{i,t}
  &=
  y_{i,t}
  -
  \hat{y}^{U,\alpha}_{i,t},
  \label{eq:si_tau_lower} \\
  \tau^{U}_{i,t}
  &=
  y_{i,t}
  -
  \hat{y}^{L,\alpha}_{i,t}.
  \label{eq:si_tau_upper}
\end{align}

A unit-time observation is classified as statistically significant if the observed value falls outside the calibrated counterfactual interval:
\begin{equation}
  \mathbbm{1}^{\mathrm{sig}}_{i,t}
  =
  \mathbbm{1}
  \left(
  \tau^{L}_{i,t} > 0
  \;\lor\;
  \tau^{U}_{i,t} < 0
  \right).
  \label{eq:si_significance_indicator}
\end{equation}
Equivalently, significance occurs when
$y_{i,t}>\hat{y}^{U,\alpha}_{i,t}$ or
$y_{i,t}<\hat{y}^{L,\alpha}_{i,t}$.

\subsection{Cumulative and relative changes}
\label{sec:si_cumulative_effect}

For each unit $i$, the cumulative deviation from expected demand up to time $T$ is:
\begin{equation}
  \Tau_{i,T}
  =
  \sum_{t=t_0}^{T}
  \tau_{i,t}.
  \label{eq:si_cum_effect}
\end{equation}

To compare changes across units with different baseline demand levels, we compute the cumulative relative change:
\begin{equation}
  \rho_{i,T}
  =
  \frac{
  \Tau_{i,T}
  }{
  \sum_{t=t_0}^{T}
  \hat{y}^{c}_{i,t}
  }.
  \label{eq:si_relative_effect}
\end{equation}
The denominator represents the cumulative counterfactual demand expected in the absence of congestion pricing.

For each unit, we also summarize the share of post-policy observations that are statistically significant:
\begin{equation}
  \pi^{\mathrm{sig}}_{i}
  =
  \frac{1}{T_{\mathrm{post}}-t_0+1}
  \sum_{t=t_0}^{T_{\mathrm{post}}}
  \mathbbm{1}^{\mathrm{sig}}_{i,t}.
  \label{eq:si_significance_share}
\end{equation}
Because a nominal $\alpha$-level prediction interval implies an expected false positive rate of $1-\alpha$, we define the excess significance share as:
\begin{equation}
  \pi^{\mathrm{excess}}_{i}
  =
  \pi^{\mathrm{sig}}_{i}
  -
  (1-\alpha).
  \label{eq:si_excess_significance}
\end{equation}

\subsection{Uncertainty aggregation}
\label{sec:si_uncertainty_aggregation}

To aggregate uncertainty over time and space, we recover the implied standard deviation of the counterfactual forecast from the calibrated prediction interval. Assuming a symmetric Gaussian approximation, the standard deviation is:
\begin{equation}
  \hat{\sigma}_{i,t}
  =
  \frac{
  \hat{y}^{U,\alpha}_{i,t}
  -
  \hat{y}^{L,\alpha}_{i,t}
  }{
  2z_{\alpha}
  },
  \label{eq:si_sigma}
\end{equation}
where
\begin{equation}
  z_{\alpha}
  =
  \Phi^{-1}
  \left(
  \frac{1+\alpha}{2}
  \right).
  \label{eq:si_z_alpha}
\end{equation}
For $\alpha=0.90$, $z_{\alpha}$ corresponds to the 95th percentile of the standard normal distribution. This Gaussian approximation is used only for aggregating uncertainty; pointwise significance is assessed directly using the calibrated prediction interval.

The cumulative variance for unit $i$ is approximated as:
\begin{equation}
  \widehat{\mathrm{Var}}
  \left(
  \Tau_{i,T}
  \right)
  =
  \sum_{t=t_0}^{T}
  \hat{\sigma}^{2}_{i,t}.
  \label{eq:si_cum_variance}
\end{equation}
The corresponding standard error is:
\begin{equation}
  \widehat{\mathrm{SE}}
  \left(
  \Tau_{i,T}
  \right)
  =
  \sqrt{
  \sum_{t=t_0}^{T}
  \hat{\sigma}^{2}_{i,t}
  }.
  \label{eq:si_cum_se}
\end{equation}
The cumulative deviation interval is:
\begin{equation}
  \left[
  \Tau^{L}_{i,T},
  \Tau^{U}_{i,T}
  \right]
  =
  \left[
  \Tau_{i,T}
  -
  z_{\alpha}
  \widehat{\mathrm{SE}}
  \left(
  \Tau_{i,T}
  \right),
  \Tau_{i,T}
  +
  z_{\alpha}
  \widehat{\mathrm{SE}}
  \left(
  \Tau_{i,T}
  \right)
  \right].
  \label{eq:si_cum_interval}
\end{equation}

The corresponding relative-change interval is obtained by scaling the cumulative deviation interval by the cumulative counterfactual baseline:
\begin{align}
  \rho^{L}_{i,T}
  &=
  \frac{
  \Tau^{L}_{i,T}
  }{
  \sum_{t=t_0}^{T}
  \hat{y}^{c}_{i,t}
  },
  \label{eq:si_relative_lower} \\
  \rho^{U}_{i,T}
  &=
  \frac{
  \Tau^{U}_{i,T}
  }{
  \sum_{t=t_0}^{T}
  \hat{y}^{c}_{i,t}
  }.
  \label{eq:si_relative_upper}
\end{align}

\subsection{Cross-unit aggregation}
\label{sec:si_cross_unit_aggregation}

To evaluate systemwide temporal dynamics, we compute the cross-unit average deviation on each day:
\begin{equation}
  \bar{\tau}_{t}
  =
  \frac{1}{N_t}
  \sum_{i=1}^{N_t}
  \tau_{i,t},
  \label{eq:si_daily_mean_effect}
\end{equation}
where $N_t$ is the number of observed units on day $t$.

The variance of the daily cross-unit mean is approximated as:
\begin{equation}
  \widehat{\mathrm{Var}}
  \left(
  \bar{\tau}_{t}
  \right)
  =
  \frac{1}{N_t^2}
  \sum_{i=1}^{N_t}
  \hat{\sigma}^{2}_{i,t}.
  \label{eq:si_daily_mean_variance}
\end{equation}

The cumulative cross-unit mean deviation is:
\begin{equation}
  \bar{\Tau}_{T}
  =
  \sum_{t=t_0}^{T}
  \bar{\tau}_{t}.
  \label{eq:si_cum_mean_effect}
\end{equation}
The cumulative variance is:
\begin{equation}
  \widehat{\mathrm{Var}}
  \left(
  \bar{\Tau}_{T}
  \right)
  =
  \sum_{t=t_0}^{T}
  \widehat{\mathrm{Var}}
  \left(
  \bar{\tau}_{t}
  \right).
  \label{eq:si_cum_mean_variance}
\end{equation}

This aggregation assumes independent forecast errors across units. If common citywide shocks remain after calibration, the resulting standard errors should be interpreted as lower-bound uncertainty estimates.

\subsection{Empirical implementation}
\label{sec:si_implementation}

We implement the framework across three demand datasets: MTA bus ridership, MTA subway ridership, and Replica-based aggregate trip counts. Bus demand is modeled at the route level using daily ridership. Subway demand is modeled at the station level using daily entries and exits. Replica-based overall travel demand is modeled at the census-tract level using weekly trip counts.

For each dataset, the base foundation model generates no-policy forecasts for the post-policy period beginning January~5,~2025. These raw forecasts are calibrated using HQC based only on pre-policy validation residuals. The calibrated forecasts are then used to compute pointwise deviations, cumulative deviations, relative changes, and significance shares.

To study spatial heterogeneity, we classify spatial units according to their relationship with the Congestion Relief Zone (CRZ). Bus routes are categorized as fully inside, partially inside, or fully outside the CRZ. Subway stations and census tracts are categorized as inside or outside the CRZ. For each stratum, we aggregate deviations from expected demand and compare relative changes across the CRZ boundary.

We use TimesFM as the main base model and Chronos as a robustness check. We also compare the proposed HQC approach with alternative calibration strategies, including raw foundation model forecasts, global quantile regression calibration, and fully per-unit quantile regression calibration. Out-of-sample validation is used to assess whether calibration improves generalization rather than overfitting the validation window.

% ======================================================================

\subsection{Model performance on validation set}
\label{sec:validation}

Table~\ref{tab:validation} summarizes the pre-policy validation performance across different forecasting and calibration methods. Overall, foundation-model-based approaches outperform traditional statistical and deep learning baselines across most datasets and evaluation metrics. In particular, calibration substantially improves both point accuracy and interval reliability for both Chronos and TimesFM. The proposed HQC framework consistently reduces RMSE and MAE relative to the uncalibrated forecasts and the global QR calibration baseline, especially for the subway and Replica datasets. Moreover, the calibrated prediction intervals achieve empirical coverage rates close to the nominal 90\% level, indicating reliable uncertainty quantification. Across the three datasets, TimesFM-HQC achieves the best overall balance between predictive accuracy and interval calibration, and is therefore adopted as the primary counterfactual forecasting framework in the subsequent analysis.

\begin{table}[!ht]
\centering
\scriptsize
\caption{\textbf{Pre-policy validation: prediction accuracy and interval calibration.}
Lower RMSE, MAE, and SMAPE indicate better point accuracy; ECR$_{0.9}$ closer to 0.9 indicates better-calibrated predictive intervals. QR denotes the global quantile regression calibration method, while HQC denotes our proposed hierarchical quantile calibration approach.}
\label{tab:validation}
\setlength{\tabcolsep}{4.5pt}
\renewcommand{\arraystretch}{0.92}
\resizebox{0.8\textwidth}{!}{%
\begin{tabular}{llrrrr}
\toprule
Dataset / Resolution & Model & RMSE & MAE & SMAPE & ECR$_{0.9}$ \\
\midrule

\multicolumn{6}{l}{\textit{Bus -- Route}} \\
 & ARIMA   & 1009.798 & 835.091  & 0.244 & 0.963 \\
 & BSTS    & 797.225  & 703.248  & 0.243 & 0.966 \\
 & Prophet & 612.293  & 481.587  & 0.170 & 0.864 \\
 & NHITS   & 542.659  & 392.966  & 0.122 & 0.842 \\
 & TFT     & 1004.688 & 848.183  & 0.252 & 0.947 \\
 & Chronos & 481.971  & 334.313  & 0.103 & 0.916 \\
 & Chronos + QR Cal. & 455.396 & 317.932 & 0.104 & 0.901 \\
 & Chronos + HQC. & 451.095 & 306.715 & 0.105 & 0.883 \\
 & TimesFM & 527.091  & 375.073  & 0.113 & 0.884 \\
 & TimesFM + QR Cal. & 482.651 & 347.196 & 0.115 & 0.901 \\
 & TimesFM + HQC. & 454.202 & 318.928 & 0.113 & 0.887 \\

\midrule
\multicolumn{6}{l}{\textit{Subway -- Station}} \\
 & ARIMA   & 1728.466 & 1446.821 & 0.235 & 0.933 \\
 & BSTS    & 2692.594 & 2260.650 & 0.223 & 0.962 \\
 & Prophet & 1235.404 & 997.466  & 0.192 & 0.739 \\
 & NHITS   & 1203.702 & 955.958  & 0.166 & 0.718 \\
 & TFT     & 1362.701 & 1055.932 & 0.177 & 0.925 \\
 & Chronos & 966.074  & 715.541  & 0.126 & 0.860 \\
 & Chronos + QR Cal. & 854.016 & 608.872 & 0.105 & 0.902 \\
 & Chronos + HQC. & 793.673 & 550.641 & 0.099 & 0.889 \\
 & TimesFM & 1073.915 & 812.636 & 0.135 & 0.821 \\
 & TimesFM + QR Cal. & 864.694 & 618.934 & 0.112 & 0.902 \\
 & TimesFM + HQC. & 803.376 & 559.534 & 0.101 & 0.894 \\

\midrule
\multicolumn{6}{l}{\textit{Replica -- Census tract}} \\
 & ARIMA   & 470.940  & 392.621 & 0.034 & 0.989 \\
 & BSTS    & 1066.099 & 940.120 & 0.074 & 0.851 \\
 & Prophet & 511.069  & 431.311 & 0.040 & 0.911 \\
 & NHITS   & 481.753  & 402.590 & 0.035 & 0.760 \\
 & TFT     & 490.526  & 413.863 & 0.035 & 0.920 \\
 & Chronos & 526.965  & 437.298 & 0.038 & 0.965 \\
 & Chronos + QR Cal. & 440.748 & 363.395 & 0.036 & 0.900 \\
 & Chronos + HQC. & 349.845 & 287.364 & 0.029 & 0.900 \\
 & TimesFM & 596.770 & 501.875 & 0.042 & 0.972 \\
 & TimesFM + QR Cal. & 479.729 & 400.123 & 0.040 & 0.900 \\
 & TimesFM + HQC. & 360.325 & 297.593 & 0.031 & 0.900 \\

\bottomrule
\end{tabular}
}
\end{table}

% ======================================================================

\subsection{Spatial bias reduction through HQC calibration}
\label{sec:spatial_bias}

In addition to improving interval calibration, we further evaluate whether the proposed hybrid quantile calibration (HQC) framework reduces spatially structured forecast bias. A key concern in counterfactual forecasting for transportation systems is that prediction errors may exhibit systematic spatial patterns, where specific regions or network segments are consistently overestimated or underestimated \citep{zhuang2025mitigating}. Such spatial bias may distort downstream effect estimation and lead to misleading interpretations of policy impacts. 

\begin{figure}[!ht]
  \centering
  \includegraphics[width=\textwidth]{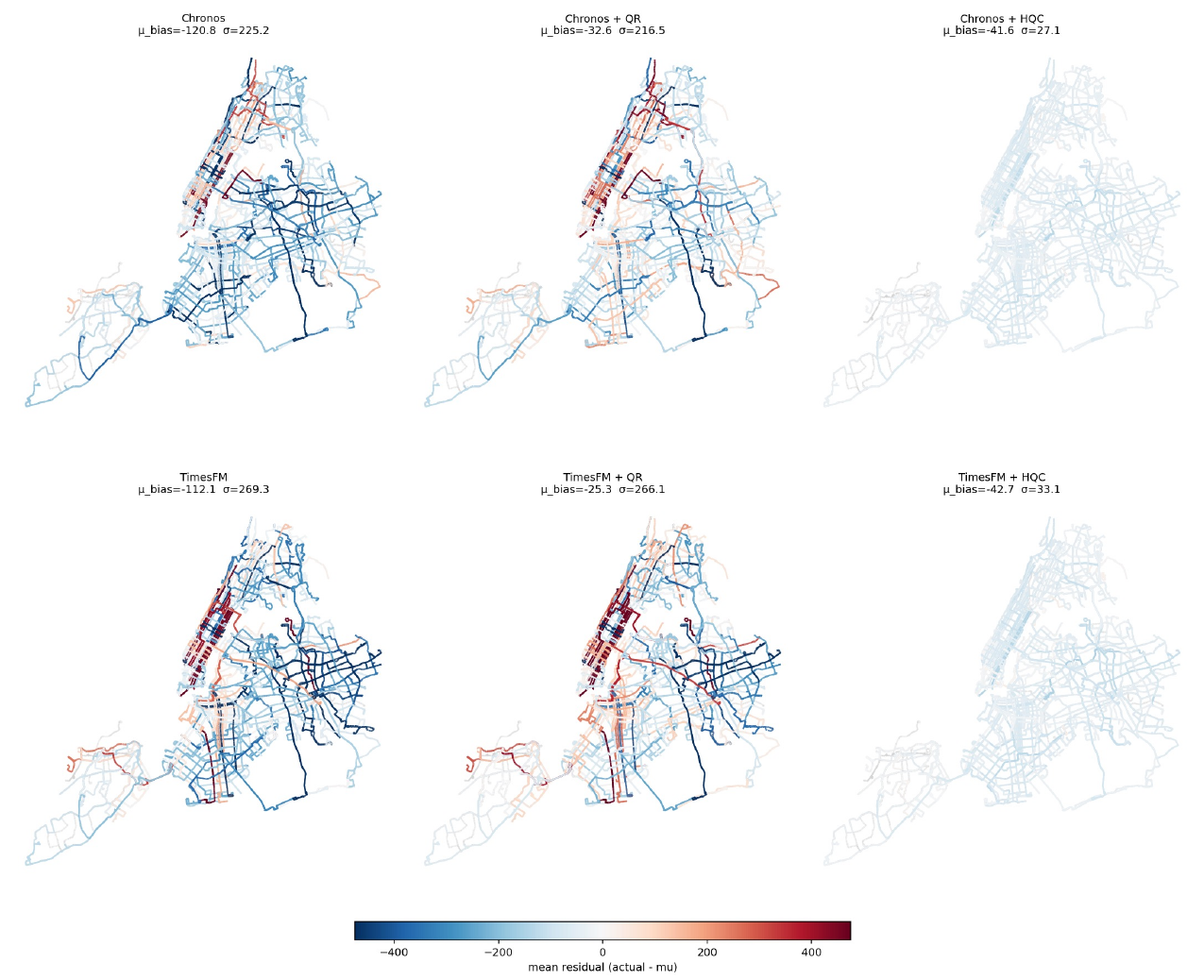}
  \caption{\textbf{Spatial distribution of residual bias for bus ridership forecasts.}
  The maps compare mean residuals for bus ridership forecasts produced by Chronos and TimesFM before calibration, after quantile regression (QR) calibration, and after the proposed hybrid quantile calibration (HQC). Compared with raw forecasts and QR-calibrated forecasts, HQC substantially reduces both the magnitude and spatial clustering of residual bias across the bus network.}
  \label{fig:spatial_bias_bus}
\end{figure}

\begin{figure}[!ht]
  \centering
  \includegraphics[width=\textwidth]{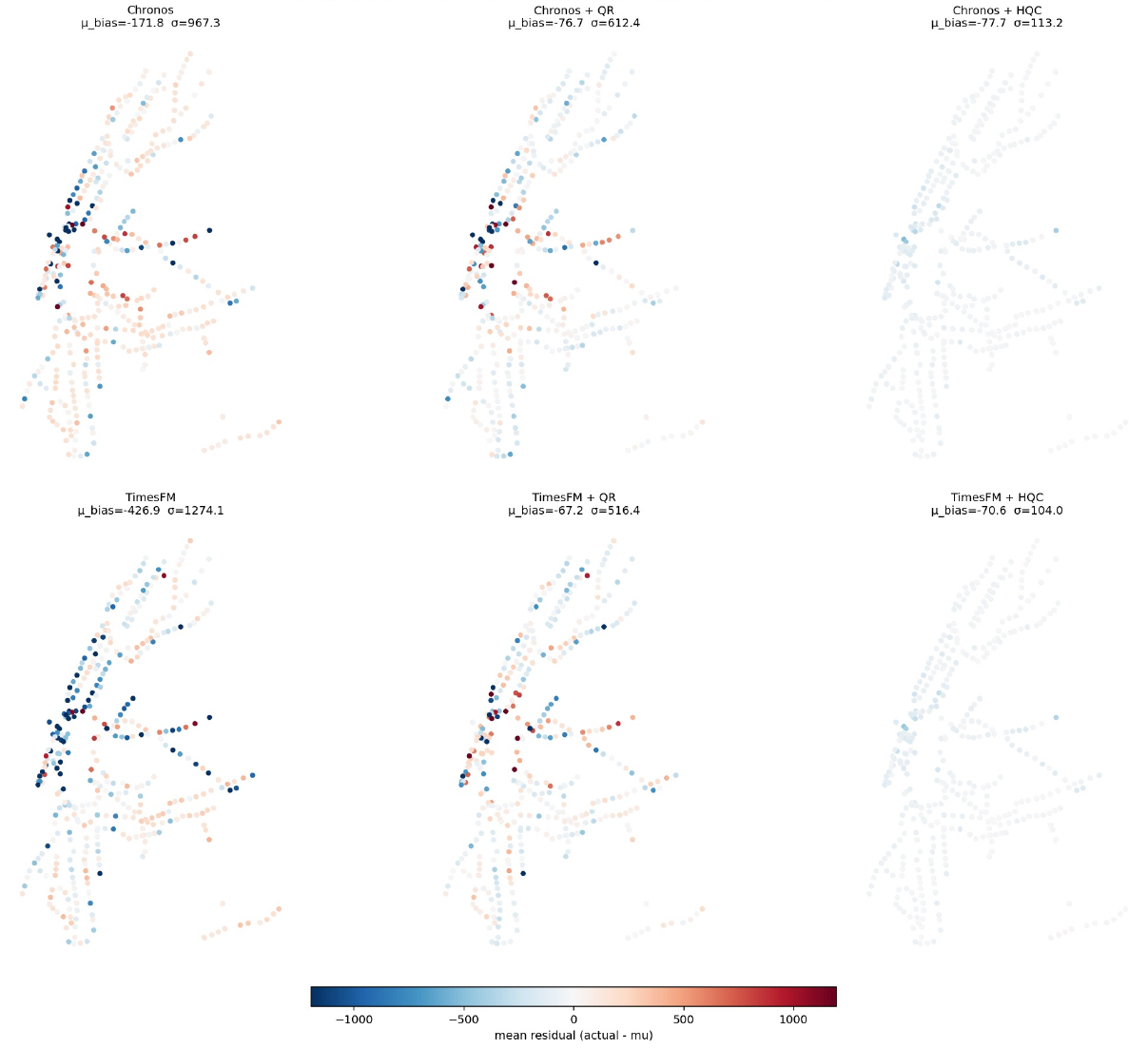}
  \caption{\textbf{Spatial distribution of residual bias for subway ridership forecasts.}
  The maps compare mean residuals for subway ridership forecasts under raw, QR-calibrated, and HQC-calibrated settings. While raw forecasts and QR calibration exhibit noticeable spatially structured errors across stations, the proposed HQC framework produces substantially more spatially balanced residual patterns with reduced bias magnitude.}
  \label{fig:spatial_bias_subway}
\end{figure}

\begin{figure}[!ht]
  \centering
  \includegraphics[width=\textwidth]{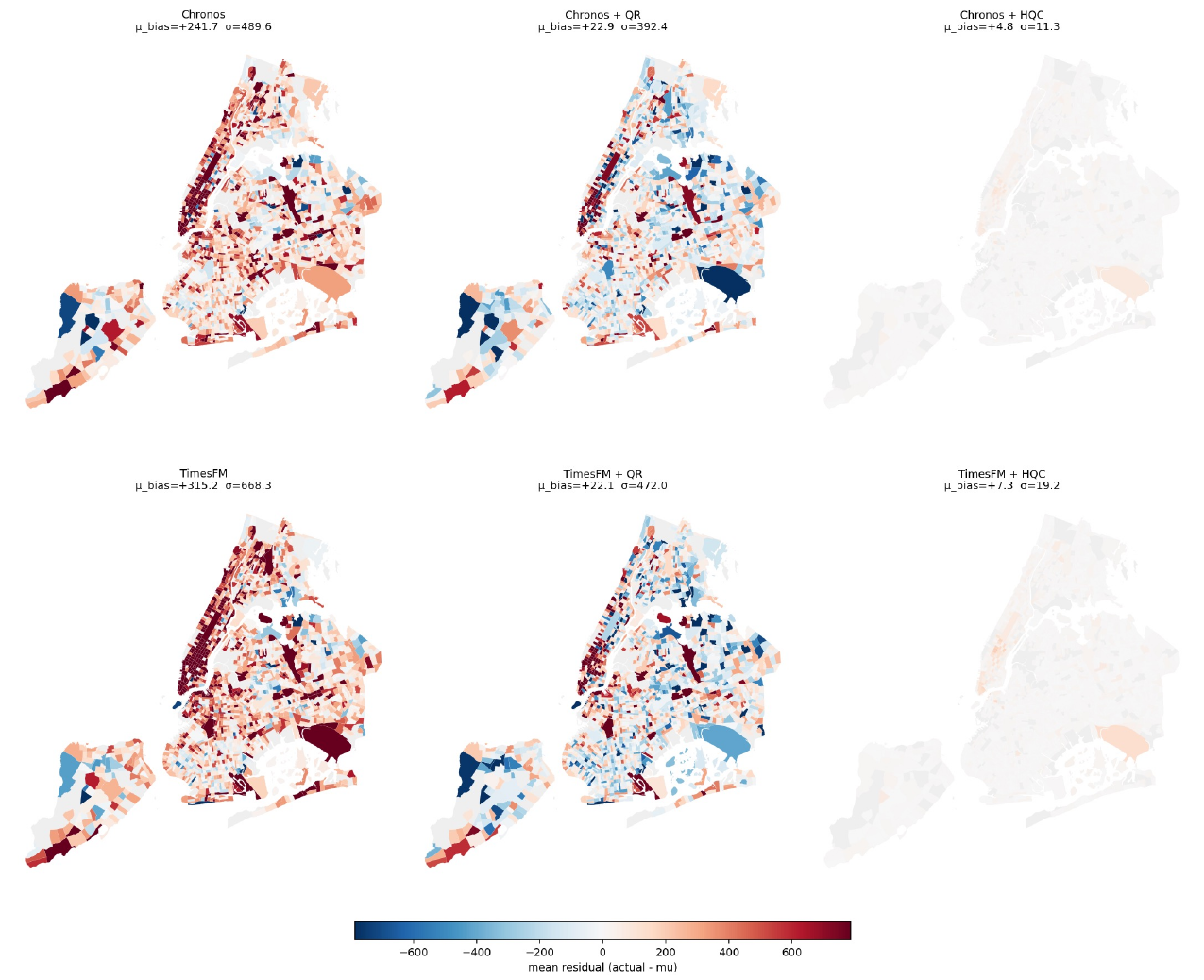}
  \caption{\textbf{Spatial distribution of residual bias for Replica mobility forecasts.}
  The maps visualize mean residuals for census-tract-level overall travel forecasts generated by different calibration methods. Compared with raw forecasts and QR calibration, the proposed HQC approach substantially reduces systematic spatial bias and produces residuals that are closer to zero across the city.}
  \label{fig:spatial_bias_replica}
\end{figure}

To examine this issue, we visualize the mean residuals (actual demand minus predicted mean demand) across spatial units for the bus, subway, and Replica datasets under three forecasting settings: raw foundation-model forecasts, quantile-regression (QR) calibrated forecasts, and the proposed HQC forecasts. Figures~\ref{fig:spatial_bias_bus}--\ref{fig:spatial_bias_replica} compare the spatial distributions of residuals for Chronos and TimesFM under these calibration strategies.

Overall, both raw forecasts and QR-calibrated forecasts exhibit clear spatially clustered residual patterns. Certain corridors and neighborhoods consistently show positive residuals, while others exhibit systematic underprediction, indicating persistent spatial bias. Although QR calibration improves probabilistic coverage, it only partially mitigates these spatially correlated errors. In contrast, the HQC framework substantially attenuates both the magnitude and the spatial clustering of residual bias across all datasets. Residual maps produced by HQC are visually closer to zero and exhibit weaker spatial structure, suggesting that the calibrated forecasts are more spatially balanced and less systematically biased.

This improvement is consistent across bus routes, subway stations, and census-tract-level Replica mobility data. The reduction in spatial bias indicates that HQC not only improves uncertainty calibration, but also enhances the reliability of spatial counterfactual estimation, which is particularly important for downstream analyses of heterogeneous effects and transportation equity.

% ======================================================================
\section{Socio-economic demographic analysis}
\label{sec:si_demographic}

\subsection{Demographic and educational covariates}

\begin{figure}[!ht]
  \centering
  \includegraphics[width=1\linewidth]{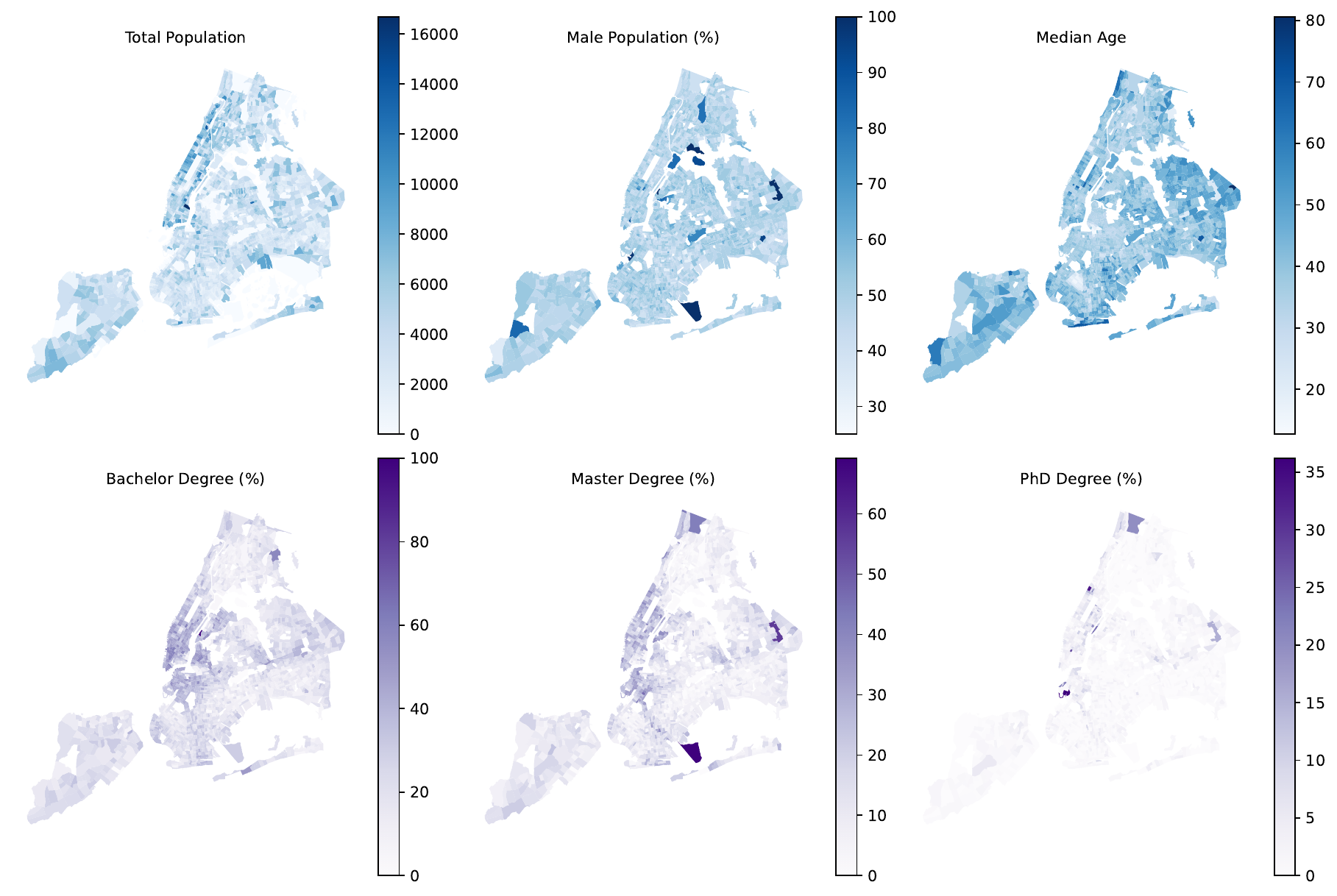}
  \caption{Spatial distribution of demographic and educational characteristics across New York City census tracts. The first row presents demographic variables (population, gender, and median age), while the second row illustrates educational attainment (share of bachelor’s, master’s, and PhD degree holders).}
  \label{fig:si_demog}
\end{figure}

Figure \ref{fig:si_demog} illustrates the spatial heterogeneity of demographic and educational characteristics across New York City census tracts. The share of male population and median age (top middle and right) display relatively balanced patterns, though older populations are slightly more prevalent in peripheral neighborhoods such as Staten Island and eastern Queens.

Educational attainment (bottom row) exhibits pronounced spatial clustering. Tracts with high proportions of bachelor’s, master’s, and PhD degree holders are predominantly located in Manhattan and certain areas of Brooklyn, aligning with the distribution of higher-income and professional households. In contrast, outer-borough tracts show lower levels of educational attainment, highlighting the educational and socioeconomic divide across the city.

\subsection{Travel-mode-share covariates}

\begin{figure}[!ht]
  \centering
  \includegraphics[width=1\linewidth]{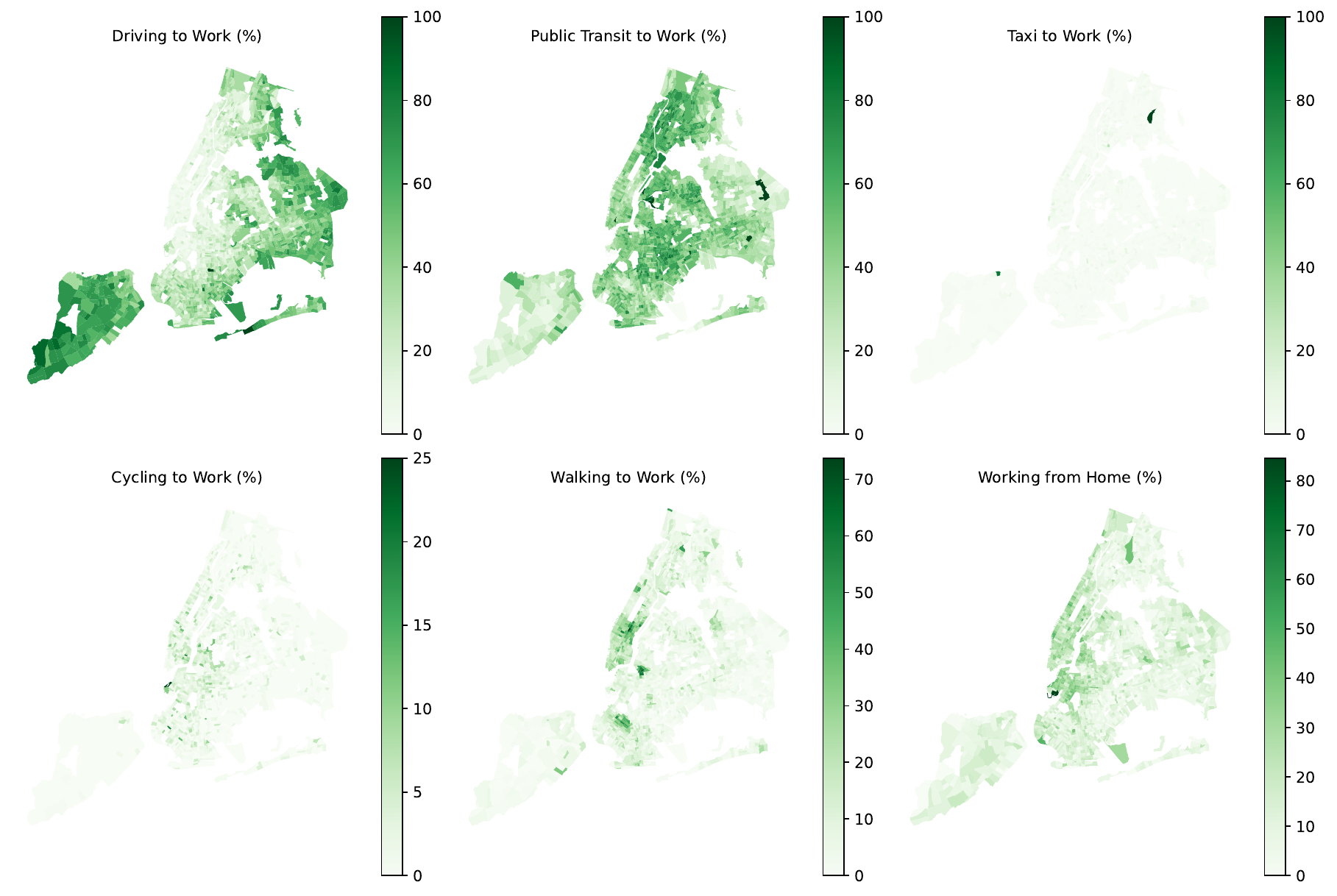}
  \caption{Spatial distribution of travel behavior in New York City. The six subplots show the shares of workers commuting by driving, public transit, taxi, cycling, walking, and working from home.}
  \label{fig:si_modes}
\end{figure}

Figure \ref{fig:si_modes} illustrates the spatial variation in commuting modes across New York City census tracts. The distribution of driving-to-work shares reveals that private automobile use is most prevalent in the outer boroughs—particularly Staten Island, eastern Queens, northeastern Bronx, and southern Brooklyn—where public transportation access is relatively limited. In contrast, reliance on public transit is concentrated in Manhattan and along major subway corridors extending into Brooklyn and the Bronx, reflecting the high connectivity of the city’s core transit network.

Non-motorized commuting modes display distinct spatial clustering. Walking and cycling to work are most common in dense, mixed-use neighborhoods such as Lower Manhattan and parts of Brooklyn near downtown areas, consistent with shorter commuting distances and higher job accessibility. Working from home shows higher shares in affluent residential tracts, especially in Manhattan and downtown Brooklyn, suggesting spatial inequalities in telework feasibility. Overall, the map highlights pronounced spatial heterogeneity in commuting behavior, closely aligned with the city’s land use structure and accessibility to transit infrastructure.

\subsection{Socio-economic covariates}

\begin{figure}[!ht]
  \centering
  \includegraphics[width=1\linewidth]{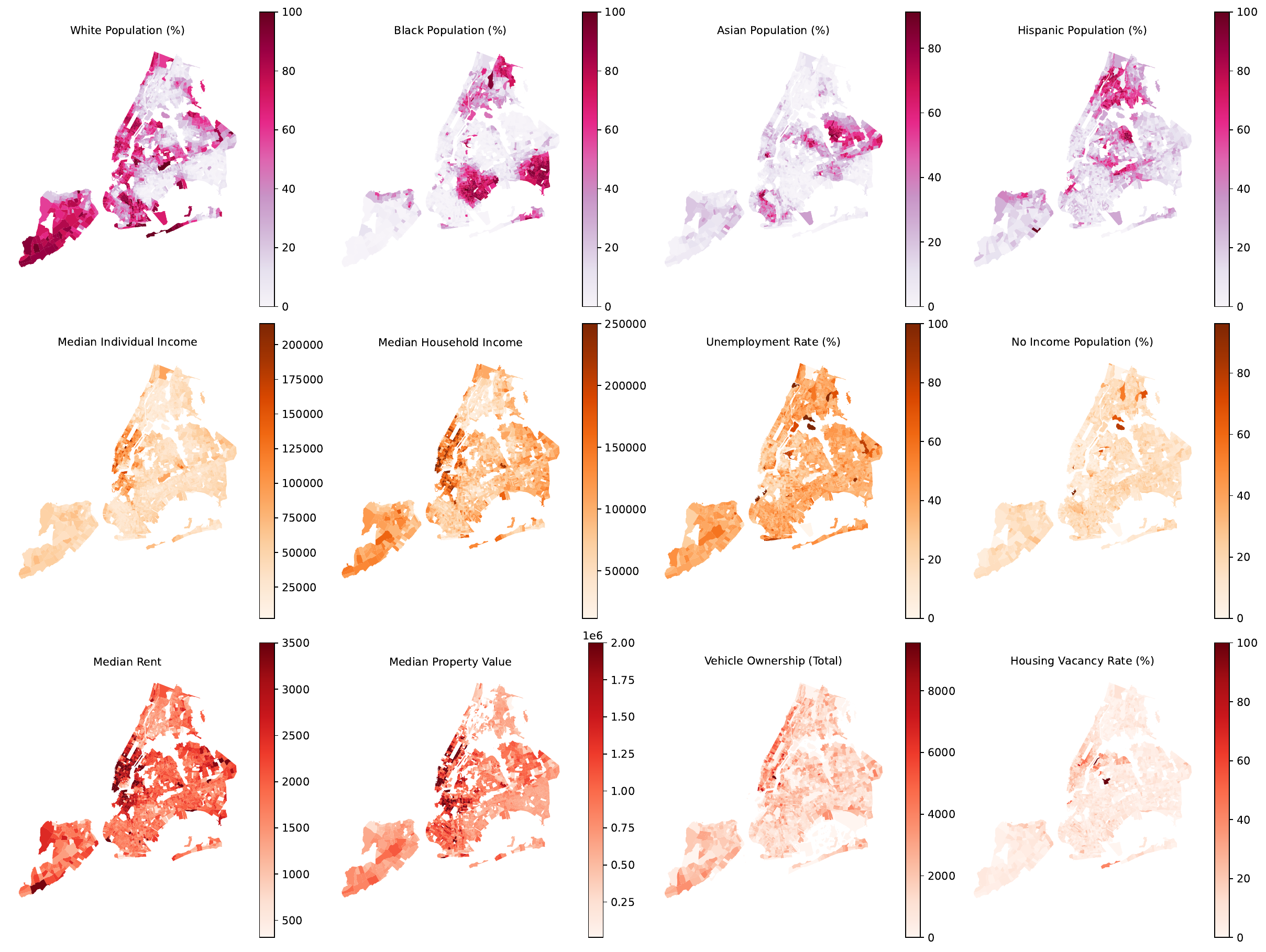}
  \caption{Spatial distribution of racial and ethnic composition, economic indicators, and housing conditions in New York City census tracts. The first row depicts racial and ethnic shares (White, Black, Asian, and Hispanic populations), the second row shows economic indicators (individual and household income, unemployment rate, and share of no-income population), and the third row presents housing and asset-related variables (median rent, property value, vehicle ownership, and housing vacancy rate).}
  \label{fig:si_socio}
\end{figure}

Figure \ref{fig:si_socio} presents the spatial distribution of racial composition, economic indicators, and housing characteristics across New York City census tracts. The racial and ethnic maps (top row) reveal clear spatial segregation patterns: White residents are predominantly concentrated in Manhattan, southern Brooklyn, and parts of western Queens, whereas Black and Hispanic populations are more prevalent in central Brooklyn, the Bronx, and southeastern Queens. Asian populations exhibit clustering in neighborhoods such as Flushing and Sunset Park, reflecting established immigrant enclaves.

Economic indicators (middle row) show similarly uneven spatial distributions. Tracts with high median individual and household incomes are concentrated in Manhattan, particularly in Midtown and the Upper East Side, as well as in portions of Brooklyn and Queens with professional or high-income populations. Areas with higher unemployment rates and larger shares of residents without income are mainly located in the Bronx and central Brooklyn, underscoring socioeconomic disparities across boroughs.

Housing and asset variables (bottom row) display strong spatial correlation with income levels. Median rent and property values are highest in Manhattan and adjacent high-income neighborhoods, while outer-borough areas exhibit lower housing costs and property values. Vehicle ownership is more common in peripheral tracts, where public transit access is limited, and vacancy rates are relatively higher in economically distressed or transitional neighborhoods. Overall, these maps highlight pronounced spatial inequalities in race, income, and housing across New York City.

% ======================================================================
\section{Spatial regression results}
\label{sec:si_spatial_reg}

We link mode-specific average daily changes relative to expected demand to U.S. Census tract
covariates (demographics, race and ethnicity, economic indicators, travel-mode shares, education and housing) and estimate maximum-likelihood spatial-lag models with a Rook contiguity spatial weights matrix. Coefficient estimates and standard errors are reported in Table~\ref{tab:si_spatial_regression}.

\begin{table}[!ht]
\centering
\scriptsize
\caption{Spatial lag regression fully standardized $\beta$ coefficients. Both $X$ and $y$ are z-scored, so each coefficient is the change in effect (in SD units of $y$) associated with a 1 SD increase in the predictor — directly comparable across modes. Significance: $^{***}p<0.001$, $^{**}p<0.01$, $^{*}p<0.05$, $^{.}p<0.1$.}
\label{tab:si_spatial_regression}
\begin{tabular}{lccc}
\toprule
Variable & Bus & Subway & Replica \\
\midrule
\textbf{Race and ethnicity} & & & \\
Share White & -0.023 & --- & 0.013 \\
 & (0.089) &  & (0.087) \\[2pt]
Share Black & -0.016 & 0.093 & -0.130 \\
 & (0.093) & (0.094) & (0.091) \\[2pt]
Share Asian & -0.028 & 0.010 & -0.033 \\
 & (0.064) & (0.080) & (0.062) \\[2pt]
Share Hispanic & 0.157$^{**}$ & 0.167$^{.}$ & -0.085 \\
 & (0.056) & (0.091) & (0.054) \\[2pt]

\textbf{Economic characteristics} & & & \\
Median individual income & --- & 0.083 & -0.092$^{*}$ \\
 &  & (0.119) & (0.038) \\[2pt]
Median household income & -0.005 & --- & --- \\
 & (0.036) &  &  \\[2pt]
Share with no income & -0.007 & -0.127 & 0.074$^{**}$ \\
 & (0.029) & (0.088) & (0.028) \\[2pt]

\textbf{Travel-mode shares} & & & \\
Share commuting by driving & -0.072$^{*}$ & 0.176$^{.}$ & 0.026 \\
 & (0.036) & (0.092) & (0.034) \\[2pt]
Share commuting by taxi & 0.064$^{**}$ & 0.089 & 0.014 \\
 & (0.023) & (0.064) & (0.022) \\[2pt]
Share commuting by cycling & -0.028 & -0.095 & 0.016 \\
 & (0.024) & (0.067) & (0.024) \\[2pt]
Share commuting by walking & -0.006 & 0.240$^{**}$ & -0.258$^{***}$ \\
 & (0.029) & (0.077) & (0.029) \\[2pt]
Share working from home & -0.026 & 0.117 & 0.101$^{**}$ \\
 & (0.036) & (0.111) & (0.035) \\[2pt]

\textbf{Education and housing} & & & \\
Share with bachelor's degree & 0.015 & 0.017 & -0.089$^{*}$ \\
 & (0.037) & (0.115) & (0.036) \\[2pt]
Share with master's degree & 0.015 & 0.028 & -0.085$^{*}$ \\
 & (0.037) & (0.117) & (0.037) \\[2pt]
Share with PhD degree & -0.003 & -0.022 & 0.087$^{***}$ \\
 & (0.027) & (0.069) & (0.026) \\[2pt]
Median property value & -0.074$^{**}$ & -0.103 & 0.001 \\
 & (0.028) & (0.082) & (0.027) \\[2pt]
Vehicles per household, imputed & -0.064$^{**}$ & -0.032 & -0.012 \\
 & (0.025) & (0.068) & (0.024) \\[2pt]
Vacancy rate & -0.047$^{.}$ & -0.007 & -0.042$^{.}$ \\
 & (0.025) & (0.075) & (0.025) \\[2pt]

\textbf{Spatial lag term} & & & \\
$W \cdot \text{Effect}$ & 0.283$^{***}$ & 0.119$^{*}$ & 0.320$^{***}$ \\
 & (0.029) & (0.060) & (0.028) \\[2pt]

\midrule
Observations $n$ & 1,771 & 271 & 1,659 \\
Pseudo $R^2$ & 0.194 & 0.106 & 0.276 \\
AIC & 4,725.1 & 776.6 & 4,261.4 \\
\bottomrule
\end{tabular}
\end{table}

% ======================================================================
\section{Robustness Check}

\subsection{Robustness to out-of-sample calibration}

To further examine the robustness of our probabilistic counterfactual framework, we repeat the calibration procedure using an out-of-sample validation strategy. This exercise evaluates whether the improvements from quantile-regression calibration are sensitive to the particular calibration sample used in the main analysis. Table~\ref{tab:robustness} reports the resulting pre-policy validation performance across bus, subway, and Replica outcomes.

The results show that the main conclusions remain stable under out-of-sample calibration. For bus ridership, both Chronos and TimesFM achieve strong predictive performance after calibration, with the intercept-adjusted versions yielding the lowest RMSE and MAE among the tested specifications. For subway demand, the calibrated foundation models substantially outperform classical baselines and neural forecasting baselines, and their empirical coverage rates remain close to the nominal 0.9 level. For Replica-based overall travel, the calibrated models also reduce point prediction errors relative to the uncalibrated foundation models.

Overall, the out-of-sample calibration results support the robustness of our framework. The relative ranking of models, the gains from calibration, and the broad calibration behavior are consistent with the main validation results. This indicates that the estimated deviations from expected demand are unlikely to be an artifact of a specific calibration design.

\begin{table}[!ht]
\centering
\scriptsize
\caption{
\textbf{Robustness check using out-of-sample calibration.}
This table reports prediction accuracy and interval calibration on the pre-policy validation period under an out-of-sample calibration strategy. Lower RMSE, MAE, and SMAPE indicate better point accuracy, while ECR$_{0.9}$ closer to 0.9 indicates better calibrated 90\% predictive intervals. Across bus, subway, and Replica outcomes, the calibrated Chronos and TimesFM models consistently improve point accuracy relative to their uncalibrated counterparts, and their empirical coverage rates remain broadly close to the nominal 0.9 level. These results suggest that the proposed counterfactual framework is robust to using an out-of-sample calibration procedure rather than relying on in-sample calibration.
}
\label{tab:robustness}
\setlength{\tabcolsep}{4.5pt}
\renewcommand{\arraystretch}{0.92}
\resizebox{0.8\textwidth}{!}{%
\begin{tabular}{llrrrr}
\toprule
Dataset / Resolution & Model & RMSE & MAE & SMAPE & ECR$_{0.9}$ \\
\midrule

\multicolumn{6}{l}{\textit{Bus -- Route}} \\
 & ARIMA   & 1009.798 & 835.091 & 0.244 & 0.963 \\
 & BSTS    & 797.225  & 703.248 & 0.243 & 0.966 \\
 & Prophet & 612.293  & 481.587 & 0.170 & 0.864 \\
 & NHITS   & 542.659  & 392.966 & 0.122 & 0.842 \\
 & TFT     & 1004.688 & 848.183 & 0.252 & 0.947 \\
 & Chronos & 481.971  & 334.313 & 0.103 & 0.916 \\
 & Chronos + QR Cal. & 477.378 & 332.842 & 0.110 & 0.885 \\
 & Chronos + HQC. & 451.095 & 306.715 & 0.105 & 0.883 \\
 & TimesFM & 527.091 & 375.073 & 0.113 & 0.884 \\
 & TimesFM + QR Cal. & 494.872 & 357.839 & 0.119 & 0.890 \\
 & TimesFM + HQC. & 454.202 & 318.928 & 0.113 & 0.887 \\

\midrule
\multicolumn{6}{l}{\textit{Subway -- Station}} \\
 & ARIMA   & 1728.466 & 1446.821 & 0.235 & 0.933 \\
 & BSTS    & 2692.594 & 2260.650 & 0.223 & 0.962 \\
 & Prophet & 1235.404 & 997.466  & 0.192 & 0.739 \\
 & NHITS   & 1203.702 & 955.958  & 0.166 & 0.718 \\
 & TFT     & 1362.701 & 1055.932 & 0.177 & 0.925 \\
 & Chronos & 966.074  & 715.541  & 0.126 & 0.860 \\
 & Chronos + QR Cal. & 871.470 & 624.210 & 0.109 & 0.894 \\
 & Chronos + HQC. & 793.673 & 550.641 & 0.099 & 0.889 \\
 & TimesFM & 1073.915 & 812.636 & 0.135 & 0.821 \\
 & TimesFM + QR Cal. & 880.159 & 632.517 & 0.114 & 0.898 \\
 & TimesFM + HQC. & 803.376 & 559.534 & 0.101 & 0.894 \\

\midrule
\multicolumn{6}{l}{\textit{Replica -- Census tract}} \\
 & ARIMA   & 470.940  & 392.621 & 0.034 & 0.989 \\
 & BSTS    & 1066.099 & 940.120 & 0.074 & 0.851 \\
 & Prophet & 511.069  & 431.311 & 0.040 & 0.911 \\
 & NHITS   & 481.753  & 402.590 & 0.035 & 0.760 \\
 & TFT     & 490.526  & 413.863 & 0.035 & 0.920 \\
 & Chronos & 526.965  & 437.298 & 0.038 & 0.965 \\
 & Chronos + QR Cal. & 458.091 & 377.633 & 0.038 & 0.880 \\
 & Chronos + HQC. & 390.169 & 321.012 & 0.034 & 0.855 \\
 & TimesFM & 596.770 & 501.875 & 0.042 & 0.972 \\
 & TimesFM + QR Cal. & 497.766 & 414.347 & 0.042 & 0.882 \\
 & TimesFM + HQC. & 402.357 & 332.468 & 0.036 & 0.857 \\

\bottomrule
\end{tabular}
}
\end{table}

\subsection{Robustness to the choice of foundation model}

\begin{figure}[!t]
    \centering
    \includegraphics[width=1\linewidth]{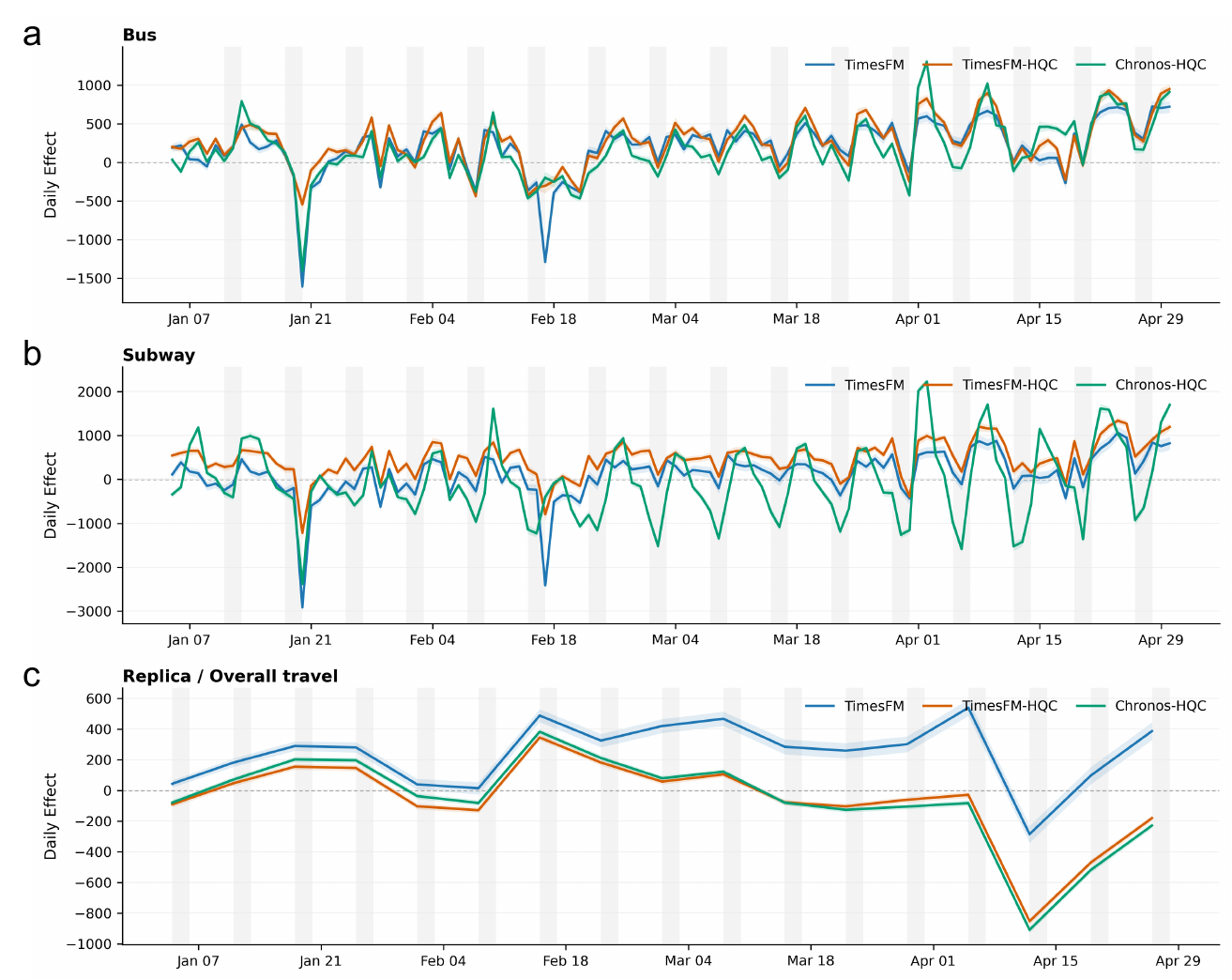}
    \caption{
    \textbf{Robustness of estimated temporal demand changes to the choice of foundation model.}
    Daily average deviations from expected demand are compared across the baseline TimesFM estimates, the calibrated TimesFM-HQC estimates, and the calibrated Chronos-HQC estimates for three mobility outcomes.
    \textbf{a,} Bus ridership.
    \textbf{b,} Subway ridership.
    \textbf{c,} Overall travel measured using Replica.
    Shaded vertical bands indicate weekends.
    Across all three modes, the Chronos-HQC estimates closely track the TimesFM-HQC estimates in both temporal pattern and magnitude. The consistency is especially clear for the timing of short-run dips, subsequent rebounds, and late-period increases, suggesting that the estimated post-policy deviations are not driven by a specific forecasting backbone. These results provide evidence that the proposed probabilistic counterfactual framework is robust to replacing TimesFM with Chronos.
    }
    \label{fig:pic6}
\end{figure}

To assess whether our estimated deviations from expected demand depend on the specific forecasting backbone, we repeat the counterfactual analysis using Chronos in place of TimesFM while keeping the same calibration and post-policy comparison procedure. Figure~\ref{fig:pic6} compares the daily average deviation trajectories obtained from the baseline TimesFM model, the calibrated TimesFM-HQC model, and the calibrated Chronos-HQC model across bus, subway, and overall travel outcomes.

The resulting temporal patterns are highly consistent across models. For bus ridership, the Chronos-HQC estimates closely follow the calibrated TimesFM-HQC trajectory, capturing similar short-run fluctuations, mid-period variation, and the increasing positive deviations toward the end of the study period. Subway deviations show somewhat larger day-to-day variability, but the Chronos-based estimates still preserve the main temporal structure identified by TimesFM-HQC, including the timing of major declines and rebounds. For overall travel, the Chronos-HQC and TimesFM-HQC estimates are also broadly aligned, especially in identifying a relatively modest response before the sharp mid-April reduction and subsequent recovery.

This consistency suggests that the substantive conclusions of the analysis are not driven by the choice of a particular time-series foundation model. Instead, the estimated deviations remain stable when replacing TimesFM with Chronos, indicating that the proposed calibrated probabilistic counterfactual framework provides robust estimates of post-policy demand changes across transportation modes.

% ======================================================================
\FloatBarrier

% ======================================================================
% REFERENCES
% ======================================================================
\FloatBarrier
\begingroup
\renewcommand{\refname}{Supplementary References}
\putbib
\endgroup
\end{bibunit}
\end{document}